%% file: alicepreprint_CDS.tex
\documentclass[ALICE,manyauthors]{cernphprep}
\usepackage{color}
\usepackage{amsmath}
\usepackage{setspace}
\usepackage{xspace}
\usepackage{subcaption}
\usepackage{booktabs}
\usepackage{multirow}

\usepackage{hyperref} 
\usepackage[figure,table]{hypcap}

\usepackage{epstopdf}
\usepackage{cite}
\usepackage[modulo]{lineno}

\newcommand{\dd}     {\mathrm{d}}
\newcommand{\dbar}   {$\overline{\mathrm{d}}$}
\newcommand{\dedx}   {d$E$/d$x$}
\newcommand{\PbPb}   {Pb--Pb}

\newcommand{\mom}    {\mbox{\rm MeV$\kern-0.15em /\kern-0.12em c$}}
\newcommand{\gmom}   {\mbox{\rm GeV$\kern-0.15em /\kern-0.12em c$}}
\newcommand{\mass}   {\mbox{\rm GeV$\kern-0.15em /\kern-0.12em c^2$}}
\newcommand{\Mmass}  {\mbox{\rm MeV$\kern-0.15em /\kern-0.12em c^2$}}
\newcommand{\pt}     {\ensuremath{p_{\rm T}\xspace}}
\newcommand{\s}      {$\sqrt{s_{\mathrm{NN}}}$}

\newcommand{\vtwo}   {$v_{2}$}
\newcommand{\pip}    {$\pi^+$}

\newcommand{\pio}    {$\pi$}
\newcommand{\hyp}    {$^{3}_{\Lambda}\mathrm H$}
\newcommand{\antihyp}{$^{3}_{\bar{\Lambda}} \overline{\mathrm H}$}

\begin{document}

%
\begin{titlepage}
\PHnumber{176}            
\PHyear{2017}           
\PHdate{13 July}            

\title{Measurement of deuteron spectra and elliptic flow \\
in \PbPb\ collisions at \s\ = 2.76 TeV at the LHC}
\ShortTitle{Deuteron spectra and elliptic flow in \PbPb\ collisions}   

\Collaboration{ALICE Collaboration%
         \thanks{See Appendix~\ref{app:collab} for the list of collaboration
                      members}}
\ShortAuthor{ALICE Collaboration}      

This publication is dedicated to the memory of our colleague H. Oeschler who recently passed away.

\begin{abstract}
The transverse momentum (\pt) spectra and elliptic flow coefficient (\vtwo) of deuterons and anti-deuterons at mid-rapidity ($|y|<0.5$)
are measured with the ALICE detector at the LHC in \PbPb\ collisions at \s\ =~2.76~TeV. 
The measurement of the \pt\ spectra of  (anti-)deuterons is done up to 8~\gmom\ in 0-10\% centrality 
class and up to 6~\gmom\ in 10--20\% and 20--40\% centrality classes. The \vtwo\ is measured in the 
0.8~$<$~\pt~$<~$5~\gmom\  interval and in six different centrality intervals (\mbox{0--5\%}, \mbox{5--10\%}, 
\mbox{10--20\%}, \mbox{20--30\%}, \mbox{30--40\%} and \mbox{40--50\%}) using the scalar product technique. 
Measured \pio$^{\pm}$, K$^{\pm}$  and p+$\overline{\mathrm{p}}$ transverse-momentum spectra and \vtwo\ are used to predict the deuteron \pt\ spectra 
and \vtwo\ within the Blast-Wave model. 
The predictions are able to reproduce the \vtwo\ coefficient in the measured \pt\ range and the transverse-momentum spectra 
for \pt~$>$~1.8~\gmom\ within the experimental uncertainties.
The measurement of the coalescence parameter $B_2$ is performed, showing a \pt\ dependence in contrast 
with the simplest coalescence model, which fails to reproduce also the measured \vtwo\ coefficient. 
In addition, the coalescence parameter $B_2$  and the elliptic flow coefficient in the 20--40\% centrality interval 
are compared  with the  AMPT model which is able, in its version without string melting, to reproduce the measured \vtwo(\pt) and the $B_2$(\pt) trend. 

\end{abstract}
\end{titlepage}

\setcounter{page}{2}

\input{introduction.tex}
\input{detector.tex}
\input{tpctof.tex}
\input{flow.tex}
\input{comparison_bw.tex}
\input{comparison_coalescence.tex}
\input{comparison_ampt.tex}
\input{conclusion.tex}
 
\newpage

\newenvironment{acknowledgement}{\relax}{\relax}
\begin{acknowledgement}
\section*{Acknowledgements}
\input{fa_2017-06-12.tex}    
\end{acknowledgement}

\bibliographystyle{utphys}
\bibliography{bibliog.bib}

\newpage

\appendix
\section{The ALICE Collaboration}
\label{app:collab}
\input{Alice_Authorlist_2017-Jun-12.tex}  
\end{document}

%% file: introduction.tex
\section{Introduction}
\label{sec:introduction}
The study of light (anti-)nuclei produced in relativistic heavy-ion collisions allows us to investigate 
the expansion and cooling down of the hot dense medium produced in heavy-ion collisions, the Quark Gluon Plasma 
(QGP), and the hadronisation mechanism. 
Proton and deuteron \pt\ spectra measured at the LHC by A Large Ion Collider Experiment  (ALICE)~\cite{Aamodt:2008zz},
show a clear dependence on the charged particle multiplicity, which can be explained by models that 
take into account the radial expansion of the emitting particle source~\cite{Adam:2015vda}. 
To investigate different production scenarios, other observables, such as the coalescence parameter ($B_A$), which corresponds to the nucleons coalescence probability, and the elliptic flow (\vtwo) of light nuclei as a function of the transverse momentum, have been already studied at SPS, RHIC and LHC \cite{Anticic:2016ckv, Adler:2004uy, Adam:2015vda,Adamczyk:2016gfs}.
The $B_A$  
values at higher \pt\ complement the available results \cite{Adam:2015vda}. Measurements of the elliptic flow \cite{Ollitrault:1992bk} allow for the investigation of 
collective effects among produced particles.
The angular distribution of all the reconstructed charged particles with respect to the
symmetry plane $\Psi_{n}$ \cite{Voloshin:2008dg} can be expanded into a Fourier series 
\begin{equation}
E \frac{\mathrm{ d^3} N}{\mathrm{d} p^3} = \frac{1}{2\pi} \frac{\rm d^2 N}{p_\mathrm{T} \mathrm{d} p_{\rm{T}} \mathrm{d}y} \left( 1 + \sum_{n=1}^{\infty} 2 v_n \cos \left( n \left( \varphi -  \Psi_{n} \right) \right) \right),
\end{equation}
where $E$ is the energy of the particle, $\vec{p}$ the momentum, $\varphi$ the azimuthal angle, $y$ the rapidity, $\Psi_{n}$ the angle of the spatial plane of symmetry of harmonic $n$ \cite{Ollitrault:2009ie,Alver:2010gr,Qiu:2011iv} and  
\begin{equation}
v_n = \langle \cos \left( n(\varphi - \Psi_{n}) \right)\rangle .
\end{equation}
The second term of the Fourier series (\vtwo) is called elliptic flow. It is directly linked to the almond shaped overlap region of the colliding ions in non central interactions and it can be related to the hydrodynamic properties of the QGP \cite{Voloshin:1994mz}. It is thus sensitive to the system conditions in the early stages of the evolution 
of a heavy-ion collision \cite{Voloshin:2008dg}. For identified hadrons \vtwo\ gives details about the hadronization mechanism. 
The deuteron is a pn bound state, whose binding 
energy ($\sim$~2.24~MeV) is about two orders of magnitude lower than the hadronisation temperature. 
Thus if it is produced at hadronisation, it is likely that it would suffer from medium induced breakup in the 
hadronic phase. 
The \vtwo\ measurements for d and \dbar\ provide an important test for the universal scaling of the elliptic flow~\cite{Nonaka:2003ew} since it is expected to scale both with the \vtwo\ of its constituent hadrons and with the \vtwo\ of the constituent quarks. \\
Comparing the measured azimuthal anisotropy of the deuteron momentum distributions 
to the proton distributions, the STAR experiment~\cite{Adamczyk:2016gfs} observed a mass 
number scaling in the 0.3~$<$~\pt~$<$~3~\gmom\ region leading to the conclusion that the mechanism of light nuclei formation at RHIC energies is 
mainly due to the coalescence of hadrons.\\  
In this paper (anti-)deuterons transverse-momentum spectra and elliptic flow \vtwo\ measured by ALICE in \PbPb\  collisions at \s~=~2.76~TeV are presented.
The paper is organised as follows: in Section~\ref{sec:alice} 
a brief description of the ALICE detector is given and in Section~\ref{section:eventselection} 
the event and track selections
used in the present analysis are described. In Section~\ref{section:spectra_analysis} 
the different techniques used to identify deuterons and anti-deuterons are presented, together with 
the efficiency and acceptance 
corrections used for the determination of the transverse momentum spectra. 
In Section~\ref{section:flow} the technique used to evaluate 
the deuteron elliptic flow and the obtained results are described, together with the 
comparison of deuteron and lighter particles 
elliptic flow. Section \ref{section:comparisons} is devoted to the comparison of the 
measured deuteron transverse momentum spectra and elliptic flow with different theoretical 
models, namely the Blast-Wave model, which is a hydro-based model\cite{Westfall:1976fu, 
Schnedermann:1993ws,Huovinen:2001cy,Adler:2001nb}, the coalescence model~\cite{Csernai:1986qf} 
and the dynamic coalescence model implemented in the AMPT generator~\cite{Lin:2004en}. 
Finally, in Section \ref{sec:conclusions} the conclusions of this work are presented.

%% file: detector.tex
\section{The ALICE detector}
\label{sec:alice}
A detailed description of the ALICE detector can be found 
in~\cite{Abelev:2014ffa} and references therein.
For the present analysis the main sub-detectors used are the V0 detector, 
the Inner Tracking System (ITS), the Time Projection Chamber (TPC), the  
Time of Flight (TOF) and the High Momentum Particle Identification Detector (HMPID)
which are located inside a maximum 0.5~T solenoidal magnetic field. 
The V0 detector~\cite{Abbas:2013taa} is formed by two arrays of scintillation counters placed around 
the beampipe on either side of the interaction point: one covering the pseudorapidity range 
$2.8 < \eta < 5.1$~\mbox{(V0-A)} 
and the other one covering $-3.7 < \eta < -1.7$~\mbox{(V0-C)}. 
The collision centrality is estimated using the multiplicity measured in the V0 detector as detailed in Section \ref{section:eventselection}. The V0 detector is also employed 
in the the elliptic flow measurement as described in section~\ref{section:flow}. 
 
The ITS \cite{Aamodt:2010aa}, designed to provide high resolution track points in the vicinity of the interaction region, is composed of three subsystems of silicon detectors placed around the 
interaction region with a cylindrical symmetry. The Silicon Pixel Detector (SPD) is the subsystem closest 
to the beampipe and it is made of two layers of pixel detectors. The third and the fourth layers are formed 
by Silicon Drift Detectors (SDD), while the outermost two layers are equipped with double-sided Silicon 
Strip Detectors (SSD). The inner radius of the SPD, 3.9~cm, is limited by the beampipe, while the 
TPC defines the radial span of the detector to be 43~cm. The ITS covers the pseudorapidity range 
$|\eta |<0.9$ and it is hermetic in azimuth.

The same pseudorapidity range is covered by the TPC \cite{Alme:2010ke}, 
which is the main tracking detector, consisting of a hollow cylinder whose axis coincides 
with the nominal beam axis. The active volume, filled with a gas at atmospheric pressure, has an inner 
radius of about 85~cm, an outer radius of about 250~cm, and an overall length along the beam 
direction of 500~cm.
The gas is ionised by charged particles traversing the detector and the ionisation electrons drift, 
under the influence of a constant electric field of $\sim $ 400~V/cm, towards the endplates where their arrival 
point is measured. The trajectory of a charged particle is estimated using up to 159 
combined measurements (clusters) of drift times and radial positions of the ionisation electrons.
The charged-particle tracks are then built by combining the hits in the ITS and 
the reconstructed clusters in the TPC. 
The tracks are then back--propagated to the beampipe to locate the primary collision position
(primary vertex) with a resolution of about 100~$\mu$m in the direction transverse to the beams for 
heavy-ion collisions.
The TPC is used for particle identification through the specific energy loss (\dedx) measurement in 
the TPC gas.

The TOF system \cite{Akindinov:2013tea} covers the full azimuth for the 
pseudorapidity interval $|\eta|<0.9$. The detector is based on the Multi-gap Resistive Plate Chambers (MRPCs) technology and it is located, with a 
cylindrical symmetry, at an average distance of 380 cm from the
beam axis. The particle identification is based on the difference between the measured 
time-of-flight and its expected value, computed for each mass hypothesis from 
track momentum and length. The detector time resolution is about 80~ps. 

The HMPID detector \cite{Abelev:2014ffa} consists of seven identical 
Ring Imaging Cherenkov (RICH) modules, in proximity focusing configuration, located at 
475~cm from the beam axis. The HMPID, with its surface of about 12~m${^2}$, covers a limited 
acceptance of \mbox{$|\eta|$ $<$ 0.55} and \mbox{1.2$^{\mathrm{o}}$ $< \phi $ $< $ 58.5$^{\mathrm{o}}$}. 
A HMPID module has three independent radiators, each one consisting of a 15 mm thick layer of liquid 
C${_6}$F$_{14}$ (perfluorohexane) with a refractive index of n = 1.289 at a photon wavelength 
$\lambda$ = 1.75 nm. They are coupled to multi-wire proportional chamber based photon detectors with CsI photocathodes. The HMPID complements the particle identification capabilities provided by the TPC and TOF detectors, extending the \pt\ reach up to 4~\gmom\ for pions and kaons and up to 6~\gmom\ for protons \cite{Adam:2015kca}.

%% file: tpctof.tex
\section{Data sample}\label{section:eventselection}
The analyses presented here are based on the data collected in the year 2011.
In total, the data sample consists of nearly 40 million \PbPb\ collisions
at \s\ = 2.76~TeV after offline event selection. 
The events are collected using a trigger logic that requires the coincidence of signals on both
sides of the V0 detector (V0-A and V0-C). An online selection based on the V0 signal amplitudes
is used to enhance the sample of central and semi-central collisions through two separate 
trigger classes. The scintillator arrays have an intrinsic time resolution better
than 0.5 ns, and their timing information is used together with that from the Zero Degree Calorimeters
\cite{Abelev:2014ffa} for offline rejection of events produced by the interaction of the beams with residual
gas in the vacuum pipe.
Furthermore, in the offline selection only events with a reconstructed primary
vertex position along the $z$ direction in the fiducial region $|V_{z}| < 10$~cm are selected.

The V0 detectors are used also to determine the centrality of Pb-Pb collisions.
The amplitude distribution of V0 is fitted with a Glauber Monte Carlo to compute the
fraction of the hadronic cross section corresponding to a given range of amplitude.
From the Glauber Monte Carlo fit it is possible to classify events in several centrality
percentiles selecting amplitudes measured in the V0 detectors as it was shown in \cite{Aamodt:2011oai,Abelev:2013qoq}. 
The contamination from electromagnetic processes is found to be negligible for
the 80\% most central events.

The \pt\ spectra and elliptic flow of primary anti-deuterons and deuterons are measured 
at mid-rapidity ($|y|~<$~0.5). A pseudorapidity selection ($|\eta|~<~$0.8) is used
in order to analyse only those tracks in the region where ALICE is able to perform
full tracking and provide the best particle identification information.
Primary particles are those produced in the collision, including all the decay products,
except those from weak decays. The main
secondary deuteron contribution comes from the knock-out deuterons produced by
the interaction of primary particles with the material of the beampipe and of
the apparatus. This is relevant for the spectra and elliptic flow measurements for \pt~$\leq$~1.4~\gmom. 
The only known contribution to secondary deuterons and anti-deuterons
from weak decays originates from the charged three-body decay of the hypertriton
(\hyp~$\rightarrow$~d~+~p~+~$\pi^{-}$) and of the anti-hypertriton 
(\antihyp$\rightarrow$~\dbar~+~$\overline{\mathrm{p}}$~+~\pip). From the
measurement of the hypertriton production via its charged two-body decay \cite{Adam:2015yta} 
we know that this contribution is negligible.

In order to guarantee a track momentum resolution of $2\%$ in the relevant $p_{\mathrm{T}}$
range and a \dedx\ resolution of about $6\%$, selected tracks are required
to have at least 70 clusters in the TPC and two points in the ITS (out of which at least one in
the SPD). The distances of closest approach to the primary vertex in the plane perpendicular (DCA$_{xy}$) and parallel (DCA$_z$) to the beam axis for the selected tracks are determined with a resolution
better than 300 $\mu m$  \cite{Abelev:2014ffa}. In order to suppress the contribution of secondary particles only tracks with $|\mathrm{DCA}_z|\leq 1$ cm are selected.
Moreover, the $\chi^{2}$ per TPC cluster is required to be less than 4 and tracks of
weak-decay products are rejected as the deuteron is a stable nucleus.

\section{Transverse momentum spectra analyses}
\label{section:spectra_analysis}
In this paper we present deuterons spectra obtained at \pt\ higher than 4.4 \gmom\ extending significantly the transverse momentum range covered in the previous ALICE study~\cite{Adam:2015vda}. 
As in the previous analysis, the spectra are determined in the centrality 
ranges 0--10\%, 10--20\% and 20--40\% consisting of 16.5, 4.5 and 9 millions of events, respectively. 
The particle identification is mainly performed by combining the information from the TPC and the TOF detectors, enabling the spectra measurement up to \pt=6 \gmom.
In the 0--10\% centrality interval  it is also possible to further extend the measurement of the production spectra to \pt=8 \gmom\ using the HMPID detector.\\
\subsection{Particle identification}
\label{subsec:spectra-pid}
\begin{figure}[!htb]  
\begin{center}
\includegraphics[width=0.45\textwidth]{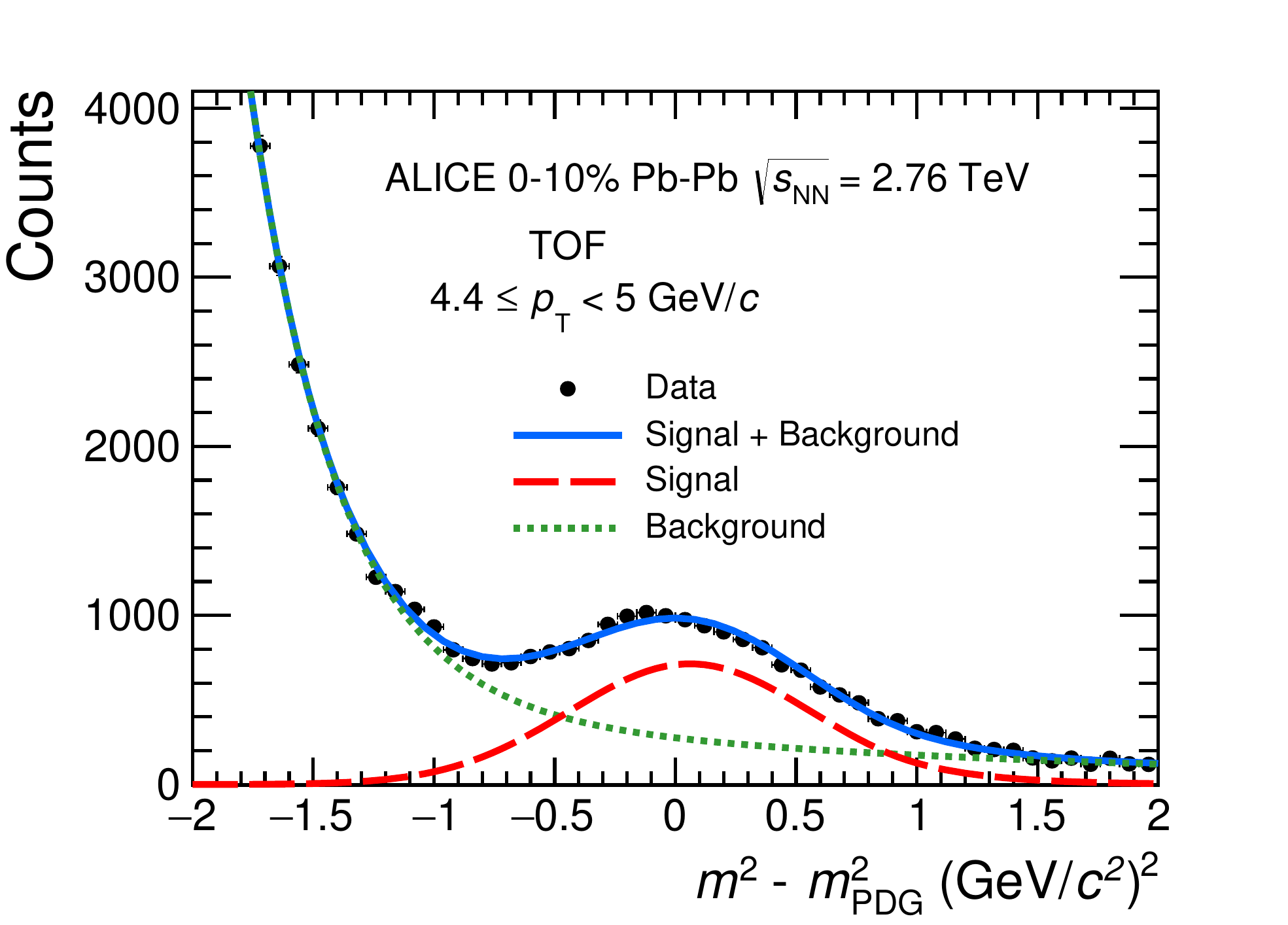} 
\includegraphics[width=0.45\textwidth]{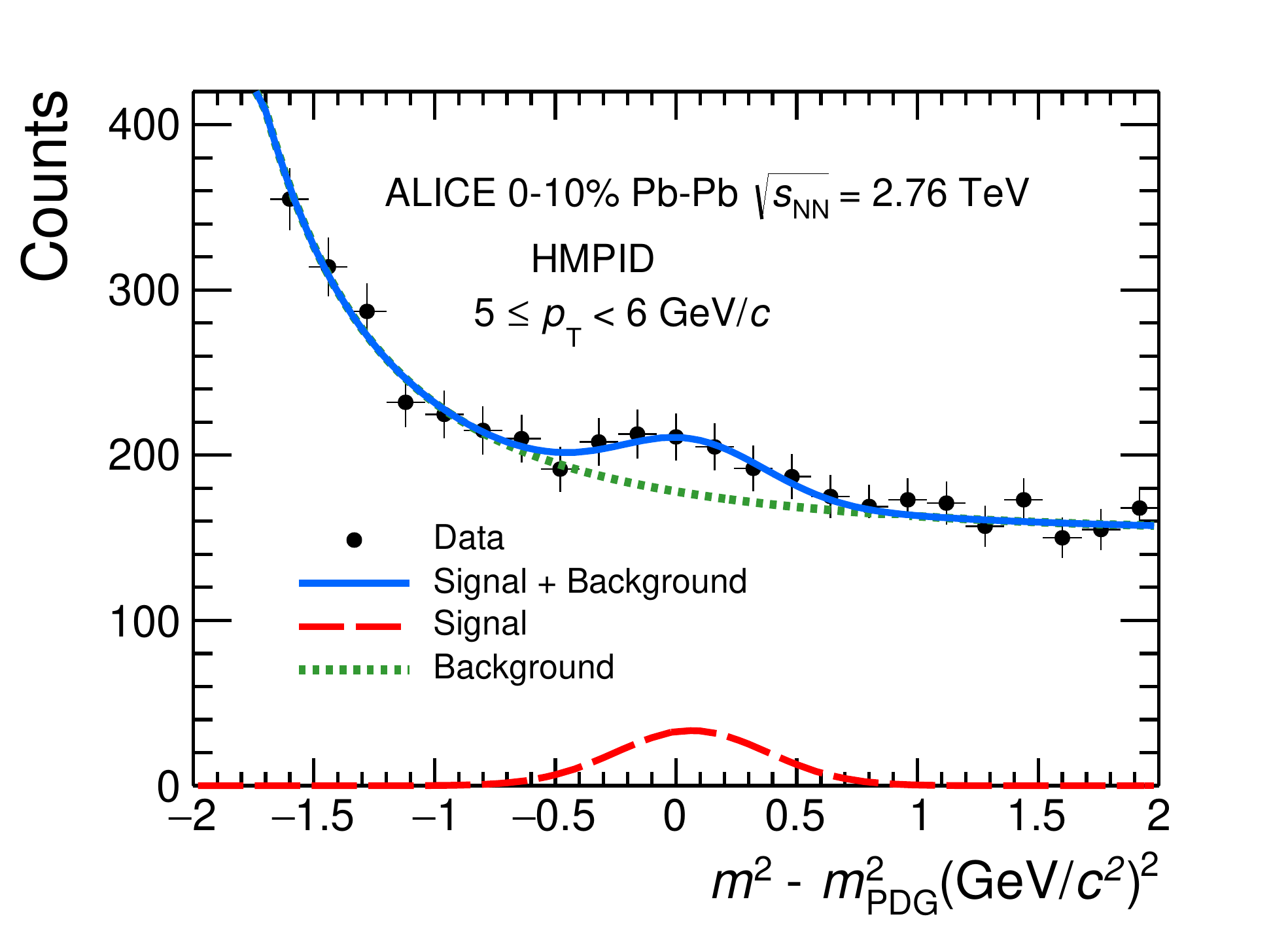} 
\end{center}
\caption{The $m^2-m_{\mathrm{PDG}}^2$ distributions obtained using the TOF detector (left) and with the HMPID detector (right) in two different \pt\  intervals 
($4.4\leq\pt<5$ \gmom\  and $5\leq\pt<6$ \gmom)  for positive tracks  in the \mbox{0--10\%} centrality class. Here $m_{\mathrm{PDG}}$ is the nominal mass of deuteron as 
reported in~\cite{Olive:2016xmw}.
Solid lines represent the total fit (signal plus background), dotted lines correspond to background and dashed lines to deuterons signal.}
\label{fig:HMPID_fit}
\end{figure}

The TPC and TOF combined analysis presented in this paper adopts the same identification strategy used in the previous ALICE measurement of light (anti-)nuclei production \cite{Adam:2015vda}.
With the large data sample collected in 2011 the deuteron transverse-momentum spectra measurement is extended up to 6 \gmom.
It is required that the measured energy-loss signal of a track as measured in the TPC lies in a $3\sigma$ window around the expected value 
for a given mass hypothesis. In addition, from the measured time-of-flight $t$ of the track, the mass $m$ of the corresponding 
particle can be obtained as:
\begin{equation}
m^{2} = \frac{p^{2}}{c^{2}}\cdot\left(\frac{c^{2}t^{2}}{L^{2}} - 1\right) .
\end{equation}
The total momentum $p$ and the track length $L$ are determined using the tracking detectors.\\
The $m^2-m_{\mathrm{PDG}}^2$ distribution, where $m_{\mathrm{PDG}}$ is the nominal mass of deuteron as 
reported in~\cite{Olive:2016xmw}, is measured for all \pt\ intervals up to 6 \gmom\ and it is fitted with a Gaussian function with an exponential tail. This is necessary to describe the asymmetric response of TOF. 
The background has two main components: the wrong association of a track with a TOF cluster and the exponential
tail of lower mass particles. For this reason the background is modelled using the sum of two exponential functions. An example of the fit used to extract the deuteron yield in the $4.4\leq\pt<5$~ \gmom\ interval for the 0--10\% centrality range is shown in the left panel of Figure \ref{fig:HMPID_fit}.

The TPC and TOF combined analysis is extended by using the HMPID measurement. With the available statistics and
due to the limited geometrical acceptance of the HMPID only results in the 0--10\% central Pb-Pb collisions are
extracted. The event and track selections are similar to those of the combined TPC and TOF analysis, but in 
addition it is required that the track is propagated to the charged-particle cluster in the MWPC of the HMPID. 
A maximal distance of 5 cm between the centroid of the charged-particle cluster and the track extrapolation on 
the cathode plane is required to reject the fake associations in the detector. This selection, tuned via 
Monte Carlo simulations, represents the best compromise between loss of statistics and the probability of an 
correct association.
The particle identification in the HMPID detector is based on the measurement of the Cherenkov angle ($\theta_{\rm{Ckov}}$)
which allows us to determine the square mass of the particle by the following formula: 
\begin{equation}
m^2 = p^2 \cdot ( n^2 \cos^2\theta_{\rm{Ckov}} - 1),
\end{equation}
where $n$ is the refractive index of the liquid radiator (C$_6$F$_{14}$ with $n$ = 1.29 at temperature 
T = 20$^{\rm{o}}$ C for photons with an energy of 6.68 eV) and $p$ is the momentum of the track. \\
In the 0--10\% centrality class, where the total number of hits in the HMPID chambers is large, the reconstruction of the Cherenkov angle is also due to photons that are not associated to the particle.
These wrong photon associations reduce the particle identification efficiency and similar effects are observed in the Monte Carlo simulations.
The response function is a Gaussian distribution
for correctly assigned rings and the raw yields are extracted by using an unfolding technique. 
The background mainly originates from fake photon associations and it is described with a 
second degree polynomial 
plus a $1/x^4$ term. Signal and background shapes are tuned
via Monte Carlo simulations, as done for lighter mass particles~\cite{Adam:2015kca}.\\
An example of the distribution of the mass squared measured with the HMPID detector in the \pt\ interval  
 $5~\leq~\pt~<6$~\gmom\  for positive tracks in the 0--10\% centrality 
interval is shown in the right part of Figure~\ref{fig:HMPID_fit}.  
Solid lines represent the total fit (signal plus background); dotted lines correspond to 
the background and dashed lines to deuterons signal.\\ 
\subsection{Corrections}
The final \pt\ spectra of (anti-)deuterons are obtained by correcting the raw
spectra for the tracking efficiency and geometrical acceptance.
The correction is defined in the same way for the two PID techniques (i.e. TPC--TOF and HMPID) 
and it is computed as the ratio of the number of detected particles to the number of generated 
particles within the relevant phase space. 
The HIJING event generator~\cite{Wang:1991h} is used to generate background events. 
To these deuterons and anti-deuterons are explicitly added with a flat distribution both in transverse 
momentum and in azimuth.
The GEANT3 transport code \cite{Brun:1994} is used to transport the 
tracks of the particles through the ALICE detector geometry. GEANT3 includes a limited simulation 
of the interaction of deuterons and anti-deuterons with the material because of the lack of 
experimental data on collisions of light nuclei with the different materials. 
For the present study, GEANT3 was modified as discussed in \cite{Adam:2015vda}: the cross-section 
of anti-nuclei are approximated in a simplified empirical model by a combination of
the anti-proton ($\sigma_{\overline{p}\mathrm{A}}$) and anti-neutron ($\sigma_{\overline{n}\mathrm{A}}$) 
cross sections, following the approach presented in ~\cite{MOISEEV1997379}. 
A full detector simulation with {\sc Geant4}~(v10.01)~\cite{Agostinelli:2002hh} has been performed in order to 
cross check the tracking efficiency estimation performed with the modified GEANT3.
Since there was a dedicated effort in the {\sc Geant4} code to interpolate the available measurements of the cross section of interaction between anti--nuclei and nuclei \cite{Galoyan:2012bh}, the correction for the interaction of (anti-)deuterons with the detector material from GEANT3 
is scaled to match the expected value from {\sc Geant4}. Half of the difference between the efficiencies 
evaluated with the two codes is 8\% for deuteron tracks
matched to the TOF, while it is 10\% for anti-deuterons tracks. This difference is taken into account in the
systematic uncertainties of the production spectra of deuterons and anti--deuterons.
The requirement of a TOF hit matched to the track reduces the overall efficiency to about 40\% 
in the \pt\ region of interest, mainly due to the TOF geometrical acceptance and to the material.

\begin{figure}[!htb]  
\begin{center}
\includegraphics[width=0.7\textwidth]{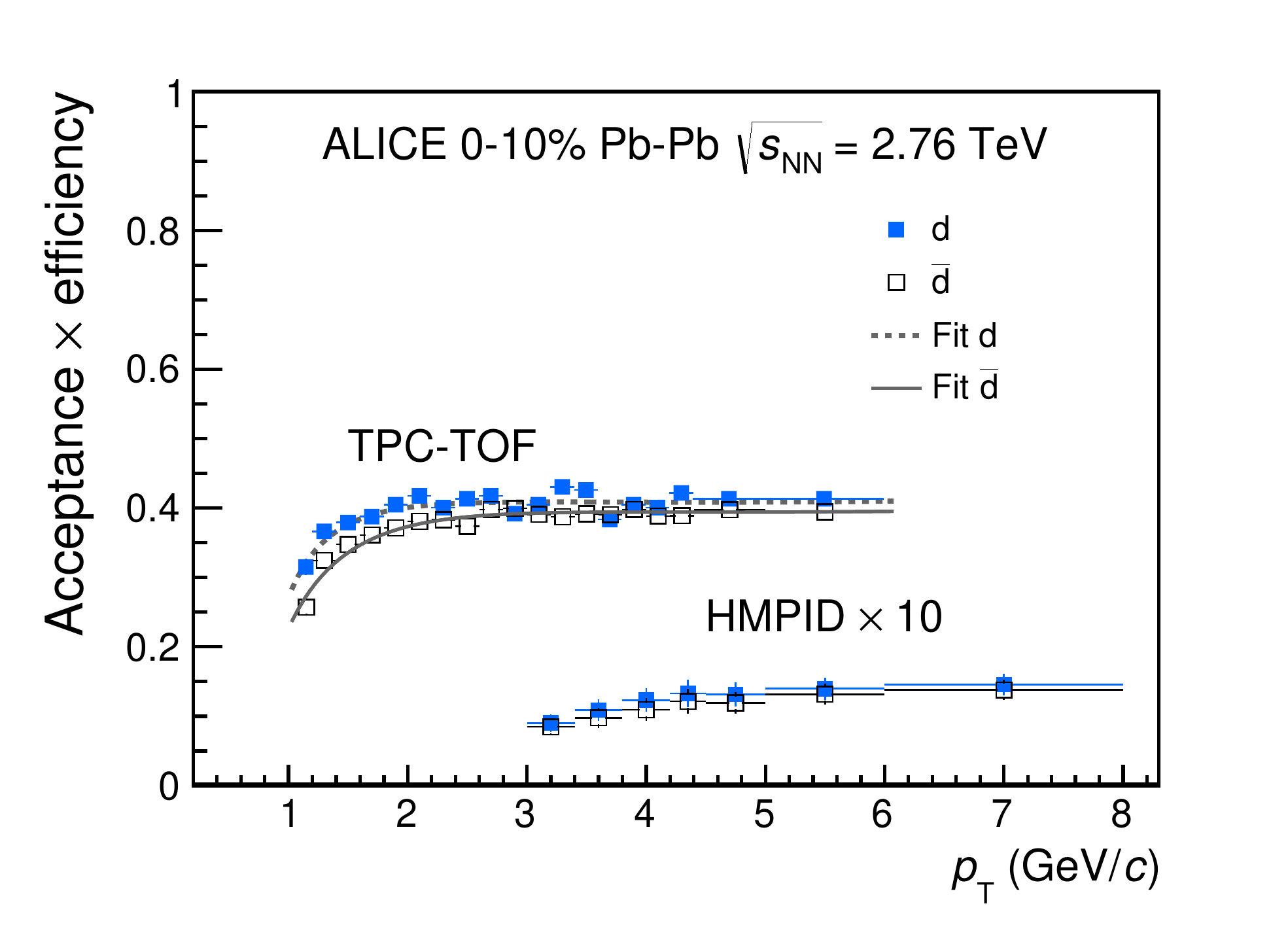}
\end{center}
\caption{Acceptance $\times$ efficiency (A $\times \epsilon$) as a function of transverse momentum for deuterons (filled markers) and anti--deuterons (open markers) in the most 10\% 
central  \PbPb\ collisions at \s\ ~=~2.76 TeV   for TPC-TOF 
and HMPID (multiplied by a scaling factor) analyses. 
The TPC-TOF points account for tracking, matching efficiency and geometrical acceptance.
The dashed and solid curves represent the fits with the function presented in equation~\ref{eq:lineTOF} for deuterons and anti--deuterons respectively (see text for details). 
The HMPID points take into account tracking efficiency,  geometrical acceptance, $\epsilon_\mathrm{{dist}}$ (`` distance correction factor" as explained in the text) and PID efficiency.  The lower value with respect to the TPC-TOF is 
mainly due to the limited geometrical acceptance of the HMPID detector (5\%).}
\label{fig:effic}
\end{figure}

Figure~\ref{fig:effic} shows the product of acceptance and efficiency (A$\times \epsilon$) for (anti-)deuterons as a function of \pt. The TPC and TOF A$\times \epsilon$ (open points) accounts for tracking efficiency, geometrical acceptance and 
matching efficiency. 
The dashed line represents a fit with the ad-hoc functional form 
\begin{equation}
f(p_\mathrm{T}) = a_0 + a_1 e^{a_2 \cdot {p_\mathrm{T}}} + a_3/p_\mathrm{T} + a_4/(p_\mathrm{T})^2,
\label{eq:lineTOF}
\end{equation}
where $a_0$, $a_1$, $a_2$ and $a_3$ are free parameters. This fit function is used to smooth the fluctuations 
in the A$\times \epsilon$ correction. However,
correcting the raw spectra with either the fit function or the binned values leads to 
negligible differences with respect to the total systematic uncertainties.
The HMPID raw spectra are corrected for tracking efficiency and geometrical acceptance 
as it  has been done for the TPC and TOF combined analysis, but the correction is higher 
mainly due to the limited geometrical acceptance of the HMPID detector. 
The HMPID particle identification efficiency is related to the Cherenkov angle reconstruction efficiency. 
It is computed by means of Monte Carlo simulations that reproduce the background observed 
in the data and it is defined as the ratio of the identified deuteron signal to the generated 
deuteron signal in the HMPID chambers. 
It reaches 50\% for (anti-)deuterons at higher transverse momenta.
A data-driven cross check of the efficiency at lower \pt\ is performed using a clean sample of (anti-)deuterons defined within 2$\sigma$ of the expected values measured by the TOF detector, showing excellent compatibility -- within statistical uncertainties -- 
between the two methods. 
In Figure~\ref{fig:effic}, the convolution of tracking efficiency, geometrical acceptance, distance correction factor ($\epsilon_\mathrm{{dist}}$) and PID efficiency for the HMPID analysis in \PbPb\ collisions at \s~=~2.76~TeV in 0--10\% centrality collisions is also shown.

The track-fitting algorithm in ALICE takes into account the Coulomb scattering and energy loss 
using the mass hypothesis of the pion. The energy loss of heavier particles, such as the deuterons, 
is considerably higher than the energy loss of pions, therefore a track--by--track correction is necessary.
This correction is obtained from the difference between the generated and the reconstructed 
momentum in a full Monte Carlo simulation of the ALICE detector.
As already discussed in \cite{Adam:2015vda}, the effect of this correction is negligible for high \pt\ 
deuterons. This momentum correction was included in systematics checks for the elliptic flow determination and 
its effect was found to be negligible. 

\subsection{Systematic uncertainties and results}
The systematic uncertainties for the two spectra analyses mainly consist of three components, in order of relevance:
\begin{itemize}
	\item transport code: the uncertainty on the hadronic cross section of the (anti-)deuterons with the material,  estimated taking the difference between the efficiencies evaluated with GEANT3 and {\sc Geant4};
	\item the fitting uncertainties for the signal extraction, studied by changing the functional form of the fitting function. The uncertainty has been estimated computing the RMS of the results of these variations;
	\item the track selection bias assessed through the variation of the track selection criteria. Among the probed selections there are the PID fiducial cut in the TPC and the track DCA$_z$ selection, whose variations turned into a negligible contribution ($\leq1\%$) to the systematic uncertainties. Since the effects of the variation of the DCA$_z$ selection are negligible, we can conclude that the production spectra of deuterons are not affected by secondary particles originating from material in the high \pt\ region.
\end{itemize}

\begin{table}[!htbp]
\begin{center}
\begin{tabular}{lcccc}
\hline
\hline
   &\multicolumn{2}{c}{TPC-TOF}        & \multicolumn{2}{c}{HMPID}    \\
  \pt\ interval (GeV$/c$)& 4.4 -- 5.0 & 5.0 -- 6.0 & 5.0 -- 6.0 & 6.0 -- 8.0 \\
  \hline
  Transport code               & \multicolumn{2}{c}{8\% (10\%)} & \multicolumn{2}{c}{8\% (10\%)}\\
  Signal extraction 		    & 3\%& 3\% & 13\% (15\%)& 15\% (18\%)\\
  Track selection  			    & 7\%& 7\% & 6\% & 7\%\\
  Material budget              &\multicolumn{2}{c}{3\%}&\multicolumn{2}{c}{3\%}\\
  HMPID $\epsilon_\mathrm{{dist}}$               &\multicolumn{2}{c}{-}  & \multicolumn{2}{c}{5\%}\\
  HMPID PID                    &\multicolumn{2}{c}{-}  & \multicolumn{2}{c}{4\%}\\
 \hline
\end{tabular}
\caption{\label{tab:syst}Details of the systematic uncertainties assigned in the TPC and TOF combined and HMPID analyses. The values in the parentheses refer to the systematic for the anti-deuteron spectra when different to the deuteron ones.}
\end{center}
\end{table}

The other contributions to the systematic uncertainties are related to the limited knowledge of the material 
budget, the PID and  the $\epsilon_\mathrm{{dist}}$ correction for the HMPID analysis. Table \ref{tab:syst} illustrates the details about 
the systematic uncertainties  for the spectra analyses in each \pt\ interval presented in this paper.

\begin{figure}[!htb]
	 \centering
     \includegraphics[width=0.5\textwidth]{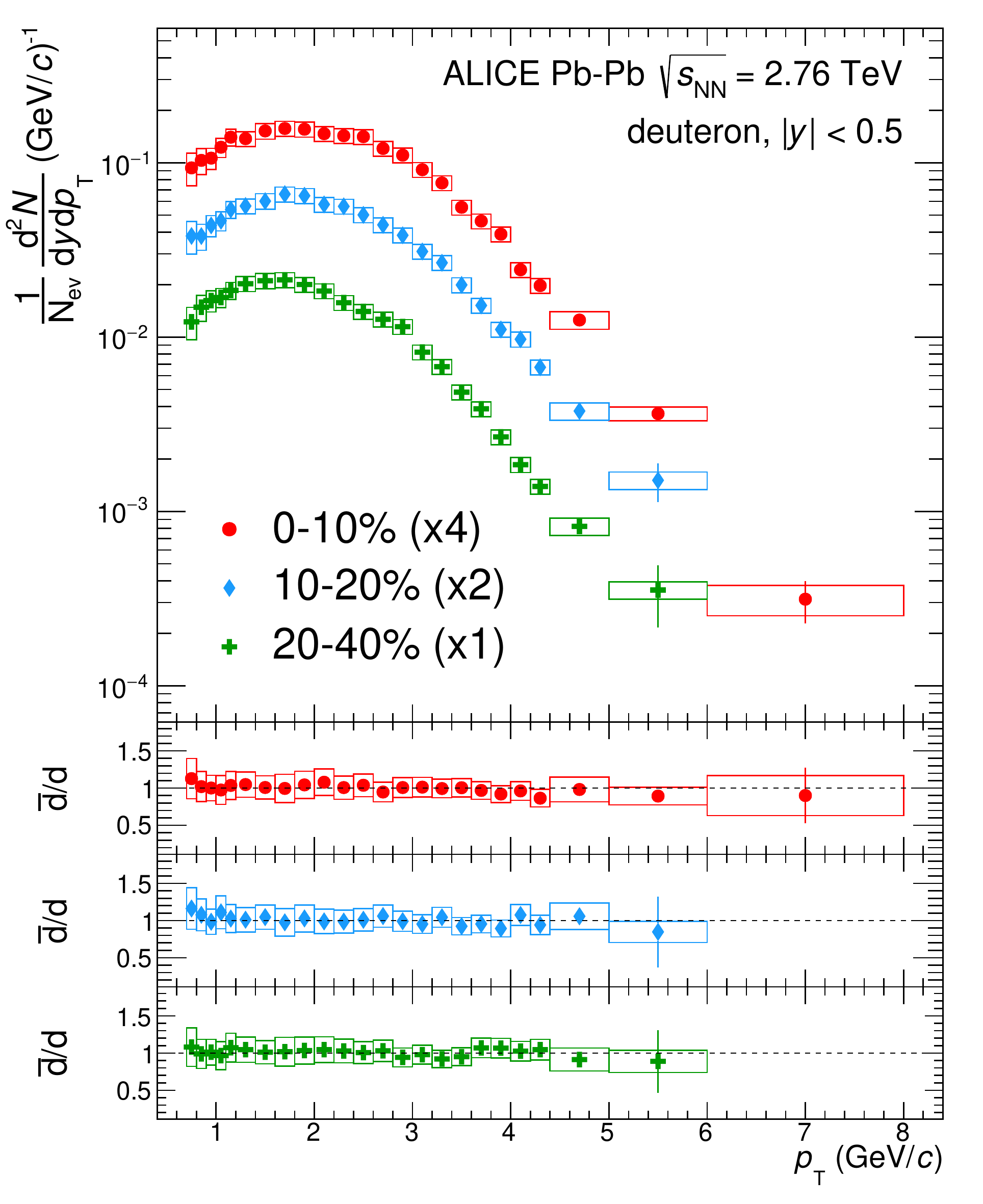}
     \caption{In the upper panel the deuteron \pt\ spectra are shown for the three centrality intervals extended to high \pt\ with the TOF and HMPID analyses.
    In the lower panels the ratios of anti-deuterons and deuterons are shown for the 0--10\%,  10--20\% and 20--40\% centrality intervals, from top to bottom.
    The ratios are consistent with unity over the whole \pt\ range covered by the presented analyses.\label{fig:ratio}}
\end{figure}

The results of the two analyses in the 0--10\% centrality interval and in the \pt\ range between 
5 and 6~\gmom\ are compatible within the uncertainties, thus in the final spectra they are 
combined using a weighting procedure. The weights used in the combination are the 
uncorrelated systematic uncertainties, given that the statistical uncertainties of the two analyses  are partially correlated. 
The resulting spectra are shown in the upper panel of Figure \ref{fig:ratio} for $\pt~>~4.4$~\gmom. For lower transverse momenta, as the data sample used for the analyses at high \pt\ presented in this paper was collected with a larger coverage of the Transition Radiation Detector and a lower performance of the Silicon Pixel Detector, the spectra extracted in \cite{Adam:2015vda} have smaller systematic uncertainties and they are used in Figure \ref{fig:ratio}. The spectra extracted with the two data samples are compatible within the systematic uncertainties.

The bottom panels of Figure \ref{fig:ratio} show the ratios between the deuteron and 
anti-deuteron spectra for the different centrality classes as a function of the transverse momentum.
As already observed in \cite{Adam:2015vda} and predicted by coalescence and thermal models the ratio is compatible with unity over the full transverse momentum region.
The integrated yield and the mean transverse momentum are extracted by fitting the spectra in each centrality interval with the Blast-Wave function \cite{Schnedermann:1993} and they are in agreement within the experimental uncertainties with the values shown in \cite{Adam:2015vda}.

%% file: flow.tex
\section{Elliptic flow measurements}
\label{section:flow}
\subsection{Analysis technique}
\begin{figure}[!htb]
\begin{tabular}{ccc}
\begin{minipage}{.5\textwidth}
\centerline{\includegraphics[width=1\textwidth]{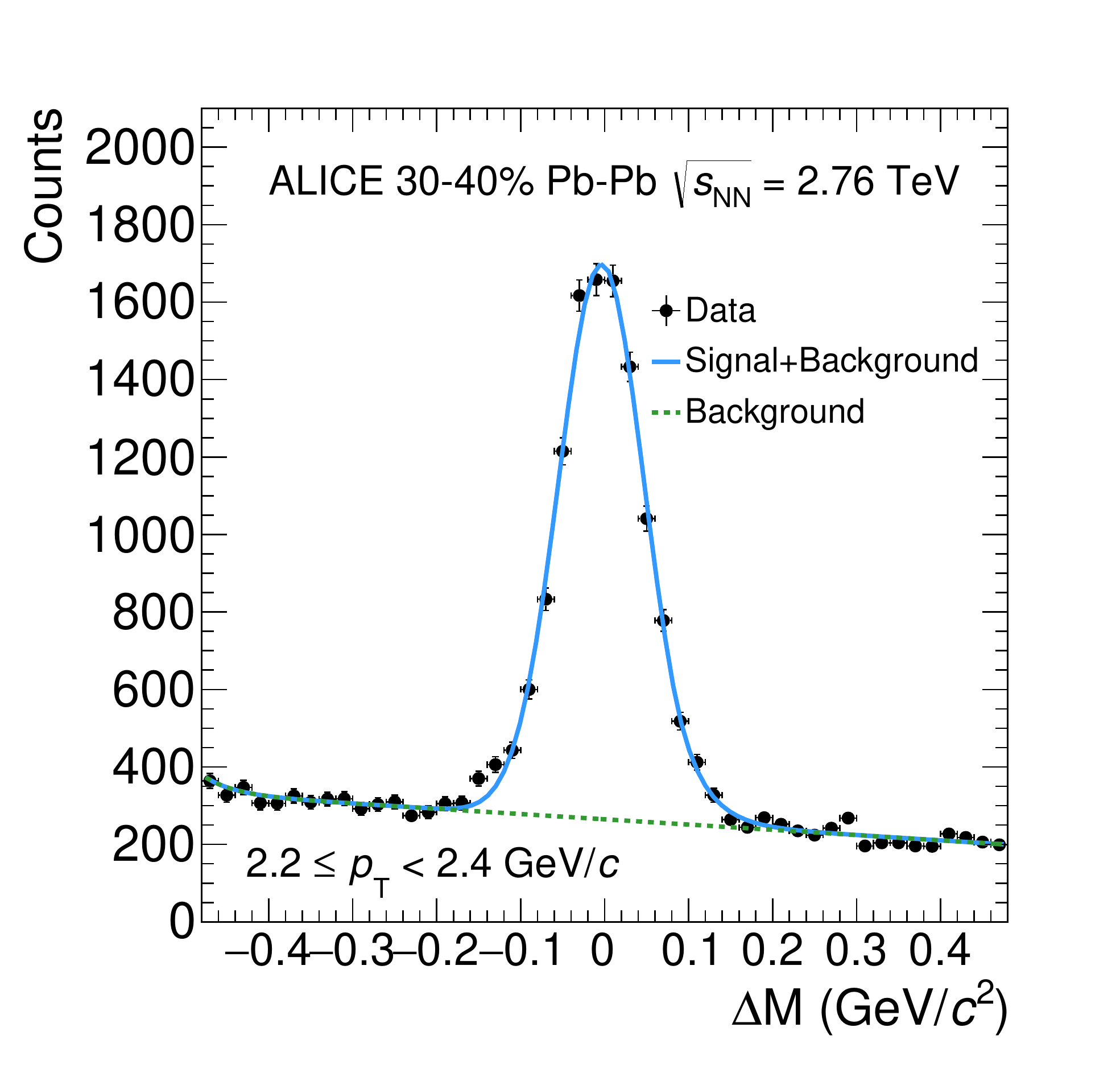}}
\end{minipage} & 
\begin{minipage}{.5\textwidth}
\centerline{\includegraphics[width=1\textwidth]{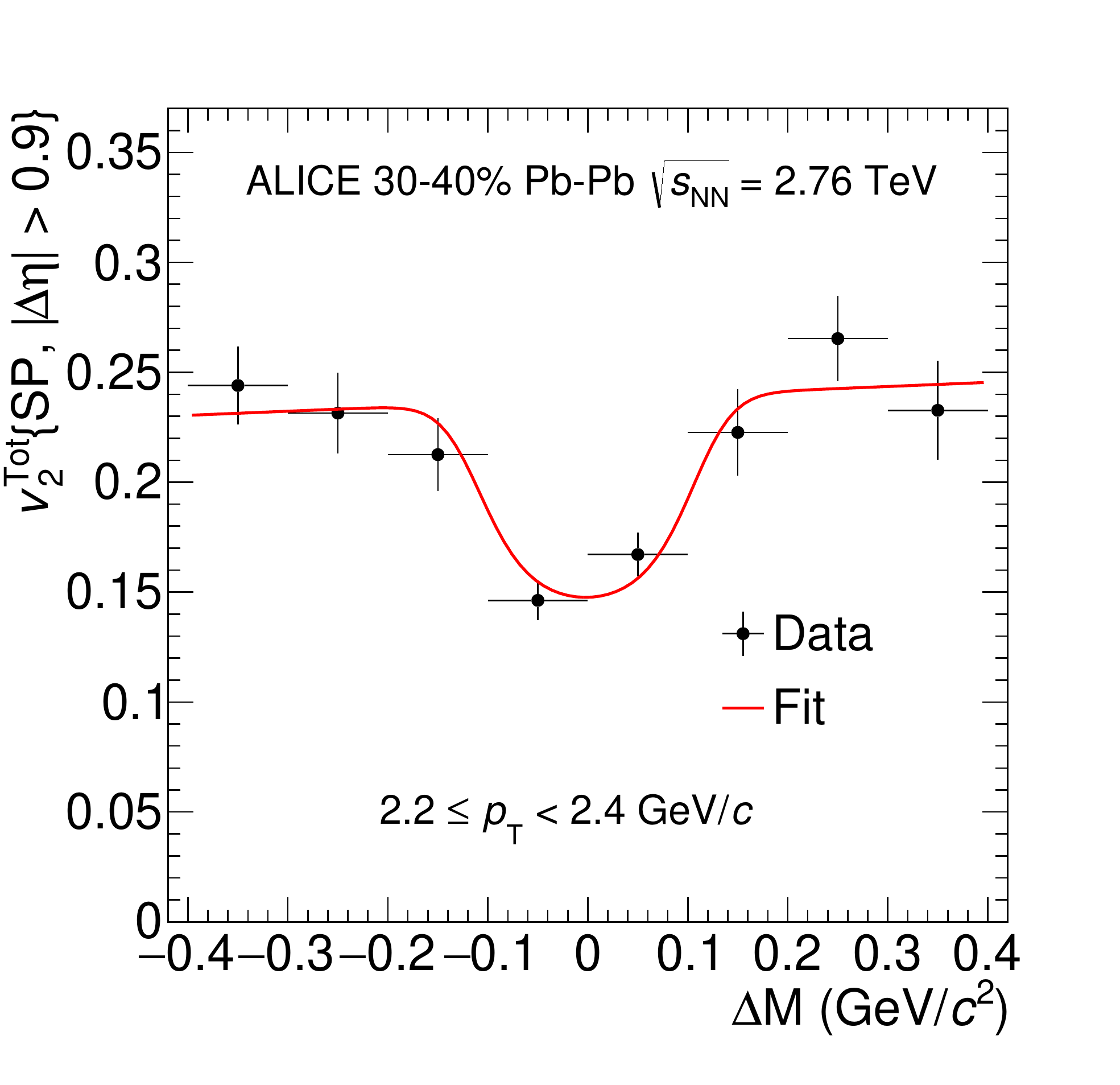}}
\end{minipage} 
\end{tabular}
\caption{Left: Distribution of $\Delta \rm{M}$ for d+\dbar\ in the $2.2~\leq$~\pt~$<2.4$~\gmom\  
and centrality interval 30-40\% fitted with a Gaussian with an exponential used to reproduce the signal and 
an exponential to reproduce the background. 
Right: The \vtwo$^{\rm{Tot}}$ vs $\Delta \rm{M}$ for d+\dbar\ for 2.2~$\leq$~\pt~$<$~2.4~\gmom\ 
in events with 30-40\% centrality. Points represent the measured $v_2^{\mathrm{Tot}}$, 
while the curve is the fit performed using equation~\ref{eq:v2tot}.}
\label{fig:method}
\end{figure}

The determination of the deuteron elliptic flow is performed over the same sample 
of \PbPb\ collisions  at \s\ = 2.76 TeV as already described in section~\ref{section:eventselection}, 
and the full event sample is divided into 6 different centrality 
intervals (\mbox{0--5\%}, \mbox{5--10\%}, \mbox{10--20\%}, \mbox{20--30\%}, 
\mbox{30--40\%} and \mbox{40--50\%}).
The identification of deuterons (d) and anti-deuterons (\dbar) is performed in the 
0.8~$<$~\pt~$<$~5~\gmom\ transverse momentum interval as 
follows: for momenta up to 1.4~\gmom\ the energy loss in the TPC gives a clean 
sample of (anti-)deuterons by requiring a maximum deviation of the specific 
energy loss of 3$\sigma$ with respect to the expected signal; above 1.4~\gmom\ 
a hit on the TOF detector is required, similarly to what has been described in the Subsection \ref{subsec:spectra-pid}.
In order to increase the statistics, deuterons and anti-deuterons are combined (d+\dbar) 
for all the centrality intervals and  in the transverse-momentum interval \pt~$>$~1.4~\gmom. 
This is possible since the results for the two separated particles are compatible within 
statistical uncertainties.  
For lower momenta (0.8~$\leq$ \pt~$<$~1.4~\gmom) only anti--deuterons are used to avoid effects 
related to secondary deuterons created through the interaction of particles with the material.   
The d+\dbar\ signal in the TOF detector is fitted with a Gaussian with an 
exponential tail, while the background is fitted with an exponential. 
An example of the $\Delta \rm{M}$ distribution, 
where \mbox{$\Delta \rm{M}$ = $m-m_{{\mathrm{PDG}}}$}, 
for deuterons plus anti-deuterons with $2.20 \leq$~\pt~$<2.40$~\gmom\ and centrality interval 
30-40\% is shown in the left part of Figure~\ref{fig:method}.\\ 
The \vtwo\ coefficient is measured using the Scalar Product (SP) method \cite{Voloshin:2008dg,Abelev:2014pua}, a two-particle 
correlation technique, using  a pseudo-rapidity gap $|\Delta \eta| > 0.9$ between the identified 
hadron under study and the reference flow particles. The applied gap reduces the non-flow 
effects (e.g. jets), which are correlations not arising from a collective motion. 
The results presented in this paper are obtained by dividing each event into three 
sub-events A, B and C, using three different pseudo-rapidity regions. 
The reference particles were taken from sub-events A and C, using the V0-A ($2.8~<~\eta~<~5.1$) and
V0-C ($-3.7~<~\eta~<~-1.7$) detectors, respectively, while deuterons were taken from sub-events 
B within $|\eta| <$0.8. 
The \vtwo\ coefficient was then calculated as described in~\cite{Abelev:2014pua}
\begin{equation} 
 v_2 = 
\sqrt{
  \frac{
  \big\langle \big\langle \vec{{u}}_2^{\mathrm{B}} \cdot \frac{\vec{Q}_2^\mathrm{A*}}{M_\mathrm{A}}\big\rangle \big\rangle
  \big\langle \big\langle \vec{{u}}_2^{\mathrm{B}} \cdot \frac{\vec{Q}_2^{\mathrm{C*}}}{M_{\mathrm{C}}}\big\rangle \big\rangle
  }{
  \big\langle \frac{\vec{Q}_2^\mathrm{A}}{M_\mathrm{A}} \cdot \frac{\vec{Q}_2^\mathrm{C*}}{M_\mathrm{C}} \big\rangle
  }
},
\label{eq:sp}
\end{equation}
\noindent where the two brackets in the numerator indicate an average over all the particles of 
interest and over all the events, $M_{\mathrm{A}}$ and $M_{\mathrm{C}}$ are the estimates of 
multiplicity from the V0-A and V0-C detectors, and $\vec{Q}_2^\mathrm{A*}$, 
$\vec{Q}_2^\mathrm{C*}$ are the complex conjugates of the flow vector \cite{Poskanzer:1998yz} calculated 
in sub-event A and C, respectively, and $\vec{{u}}_2^{\mathrm{B}}$ is the unit flow vector 
measured in sub-event B.  
The contribution to the measured elliptic flow (\vtwo$^{\mathrm{Tot}}$) due to misidentified
deuterons (\vtwo$^{\rm{Bkg}}$) is removed by studying the azimuthal correlations versus 
$\Delta \rm{M}$. This method is based on the observation 
that, since \vtwo\ is additive, candidates \vtwo$^{\rm{Tot}}$ can 
be expressed as a sum of signal (\vtwo$^{\rm{Sig}}$($\Delta$M)) and background 
(\vtwo$^{\rm{Bkg}}$($\Delta$M)) weighted by their relative yields 
\begin{equation} 
v_2^{\mathrm{Tot}} (\Delta\mathrm{M}) = v_2^{\mathrm{Sig}}(\Delta\mathrm{M}) \frac{\mathrm{N}^{\mathrm{Sig}}}{\mathrm{N}^{\mathrm{Tot}}}(\Delta\mathrm{M}) +  v_2^{\mathrm{Bkg}}(\Delta\mathrm{M}) \frac{\mathrm{N}^{\mathrm{Bkg}}}{\mathrm{N}^{\mathrm{Tot}}}(\Delta\mathrm{M}),
\label{eq:v2tot}
\end{equation}
where N$^{\rm{Tot}}$ is the total number of candidates, N$^{\rm{Bkg}}$ and N$^{\rm{Sig}}$ = N$^{\rm{Tot}}$ - N$^{\rm{Bkg}}$ 
are the numbers of background and signal for a given mass and \pt\ interval.  
The yields N$^{\rm{Sig}}$ and N$^{\rm{Bkg}}$ are extracted from fits to the 
$\Delta \rm{M}$ distributions obtained with the TOF detector for each centrality and \pt\ interval. 
The  \vtwo$^{\rm{Tot}}$ vs $\Delta \rm{M}$ for 
d+\dbar\ for 2.2~$\leq~$\pt~$<$~2.4 \gmom\ in events with 30--40\% centrality is shown in the right panel of 
Figure~\ref{fig:method}, where the points represent the measured $v_2^{\mathrm{Tot}}$ and the curve is the 
fit performed using equation~\ref{eq:v2tot}. The $v_2^{\mathrm{Bkg}}$ was parametrized 
as a first-order polynomial ($v_2^{\mathrm{Bkg}} (\Delta\rm{M}) = p_0 + p_1\ \Delta\rm{M}$).

\subsection{Systematic uncertainties and results}
The systematic uncertainties are determined by varying the event and track selections. 
The contribution of each source is estimated, for each centrality interval, as the  root mean square 
deviation of the \vtwo(\pt) extracted from the variations of the cut values relative to the results 
described above. 
The total systematic uncertainty was calculated as the quadratic sum of each individual contribution.  
The event sample is varied by changing the cut on the position of the primary 
vertex along the beam axis from $\pm$ 10 cm to $\pm$ 7 cm, by replacing the 
centrality selection criteria from the amplitude of the signal of the V0 detector to the multiplicity of the 
TPC tracks and by separating runs with positive and negative polarities of the solenoidal magnetic field. 
The systematic uncertainties related to these changes are found to be smaller than 1\%. 
Additionally, systematic uncertainties related to particle identification are studied by 
varying the number of standard deviations around the energy loss expected for deuterons in the TPC 
and, similarly, for the time of flight in the TOF detector and by varying the distance of closest 
approach in the DCA$_{xy}$ of accepted tracks. 
These contributions are found to be around 2\% for all the measured  
transverse-momentum and centrality intervals. 
The  systematic uncertainties originating from the determination of N$^{\rm{Sig}}$, N$^{\rm{Tot}}$ 
and N$^{\rm{Bkg}}$ in equation \ref{eq:v2tot}, are studied by using different functions 
to describe the signal and the background. 
The function adopted to describe the  $v_2^{\mathrm{Bkg}} (\Delta\rm{M})$ is varied using different polynomials 
of different orders. The contribution to the final systematic uncertainties is found to be 
around 3\% for all the analysed transverse-momentum and centrality intervals. 
The main contributions to the systematic uncertainties of deuteron elliptic flow are related to 
TPC and TOF occupancy~\cite{Abelev:2014pua}. 
These contributions were studied in detail in~\cite{Abelev:2014pua} and are adopted 
in the present analysis, leading to absolute systematic uncertainties of 0.02 and 0.01 related to TPC and TOF 
occupancy, respectively.  A summary of all the systematic uncertainties can be found in 
Table~\ref{table:summsystFlow}.
 
\begin{figure}[!htbp]
\begin{center}
\includegraphics[width=0.6\textwidth]{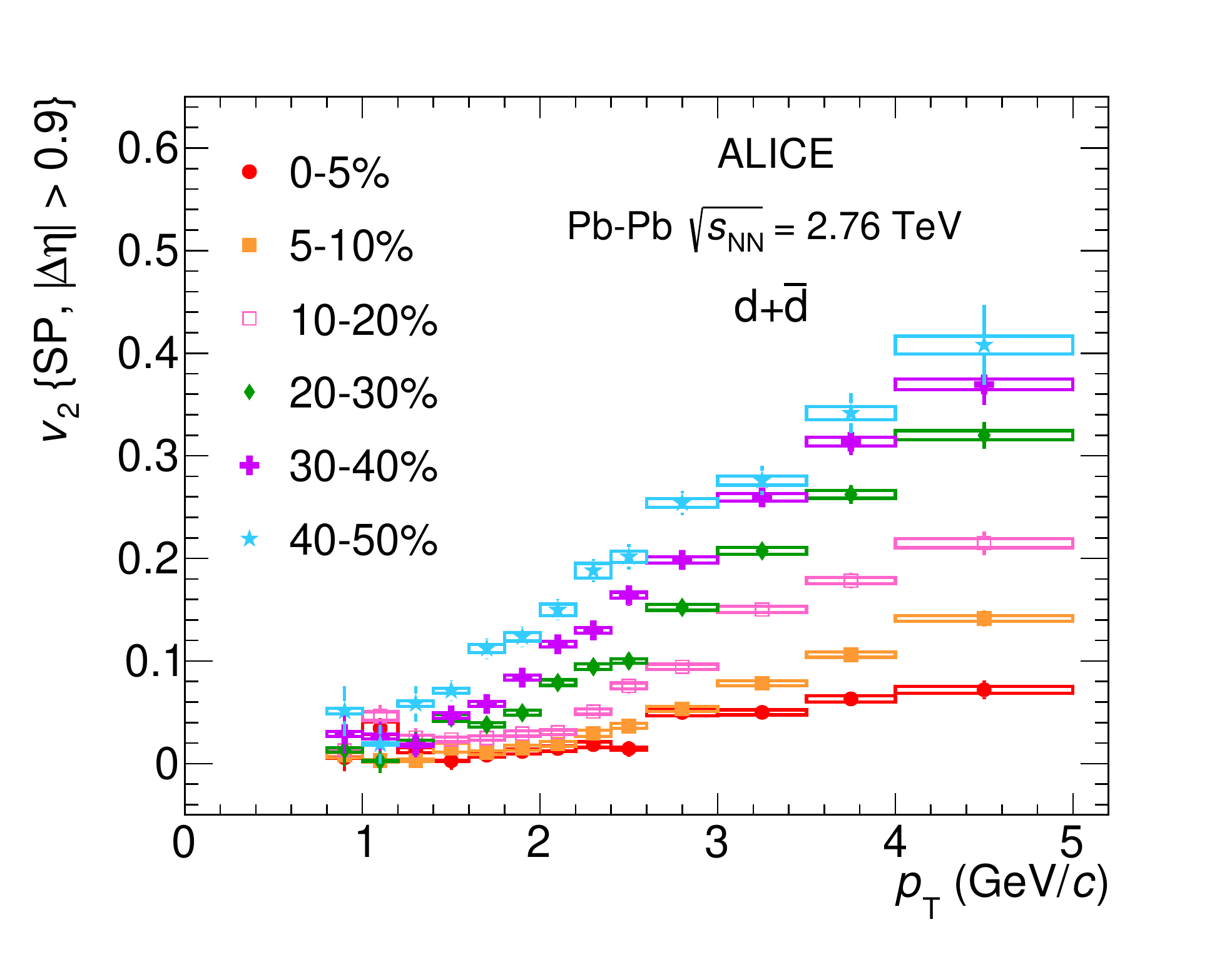}
\caption{Measured \vtwo\ as a function of \pt\ for \dbar\ (\pt~$<$~1.4~\gmom) and d+\dbar\ 
(\pt~$\geq$~1.4~\gmom) for different centrality intervals in \mbox{\PbPb} collisions 
at \s~=~2.76~TeV. Vertical bars represent statistical errors, while boxes are systematic uncertainties.}
\label{fig:Measurement}
\end{center}
\end{figure}

\begin{table}[!h]
\begin{center}
\begin{tabular}{l c }
\hline
\hline  Source                                                         & Value \\
\hline  Event Selections                                               & $<$1\%  \\
        Particle Identification                                        & 2\% \\
        Fit to \vtwo$^{\rm{Tot}}$ vs $\Delta \rm{M}$                   & 3\% \\ 
        TPC and TOF occupancy (absolute value)                         & 0.02(TPC) 0.01 (TOF)\\ 
\hline
\end{tabular} 
\caption {Summary of the systematic uncertainties for the determination of the deuterons  \vtwo\ coefficient.}
\label{table:summsystFlow}
\end{center}
\end{table}
The measured \vtwo\ as a function of \pt\ for d+\dbar\ is shown in Figure~\ref{fig:Measurement}. 
Each set of points corresponds to a different centrality class: \mbox{0--5\%}, \mbox{5--10\%}, 
\mbox{10--20\%}, \mbox{20--30\%}, \mbox{30--40\%} and \mbox{40--50\%}, as reported in the legend. 
Vertical lines represent statistical errors, while boxes are systematic uncertainties.  
The value of \vtwo(\pt) increases progressively from central to semi-central collisions. 
This behaviour is consistent with the picture of the final-state anisotropy driven by the collision 
geometry, as represented by the initial-state eccentricity which decreases from peripheral 
to central collisions.  
\subsection{Comparison with other identified particles and test of scaling properties}
\begin{figure}[!htb]
\begin{center}
\includegraphics[width=1\textwidth]{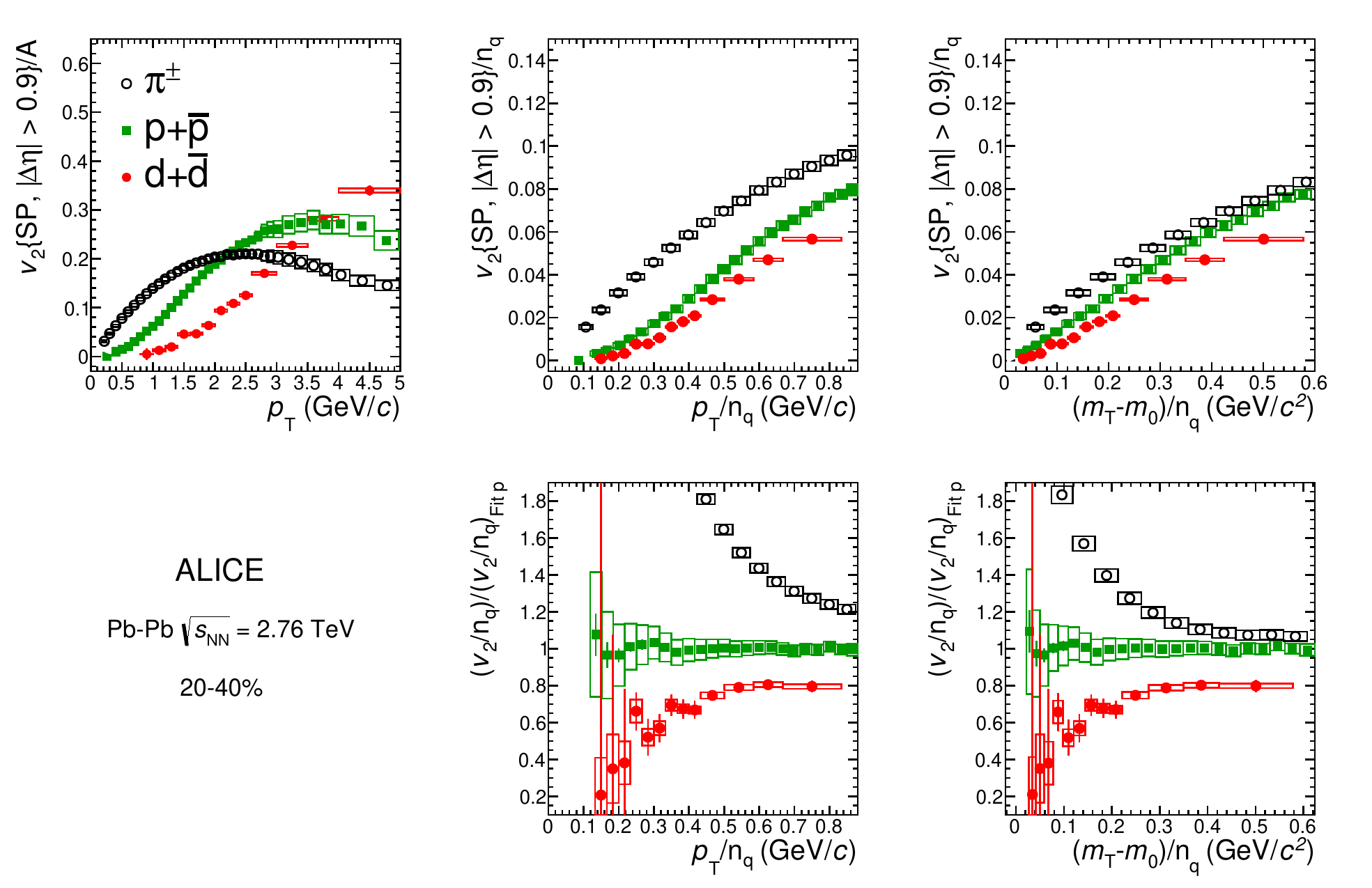}
\caption{\vtwo\ of $\pi^{\pm}$  (empty circles), p+$\overline{\rm p}$ (filled squares) and d+\dbar\ (filled circles) measured in 
the 20--40\% centrality interval. A detailed description of each panel can be found in the text.}
\label{fig:ComparisonMeasured1}
\end{center}
\end{figure}

In order to study the spectra and the elliptic flow of deuterons simultaneously, the latter has been  
determined in the same centrality intervals selected for the \pt\ spectra 
(\mbox{0 -- 10\%}, \mbox{10 -- 20\%} and \mbox{20 -- 40\%}) (see Section~\ref{section:spectra_analysis}). 
The measured \vtwo\ coefficient for d+\dbar\ is compared with that of pions and protons~\cite{Abelev:2014pua}. 
The results in the 20--40\% centrality interval are shown in Figure~\ref{fig:ComparisonMeasured1}. 
The \vtwo \ of $\pi^{\pm}$  (empty circles), p+$\overline{\rm p}$ (filled squares) and d+\dbar\ (filled circles) 
as a function of \pt\ are shown in the top left panel of the figure.
It is observed that at low \pt\ deuterons follow the  mass ordering observed for lighter particles, which is attributed 
to the interplay between elliptic and radial flow \cite{Huovinen:2001cy, Shen:2011eg}. 
The second column of Figure~\ref{fig:ComparisonMeasured1} is used to test the scaling properties of \vtwo\ with the 
number of constituent quarks (n$_{\rm q}$).   
It has been observed at RHIC \cite{Adams:2003am, Abelev:2007qg,  Adare:2006ti} that the various identified hadron species approximately show a follow a common behaviour  \cite{Afanasiev:2007tv}, while nuclei follow an atomic mass number scaling in the 0.3~$<$~\pt~$<$~3~\gmom\ interval~\cite{Adamczyk:2016gfs}.
The \vtwo\ coefficient divided by n$_{\rm q}$  is shown as a function of  \pt/n$_{\rm q}$ in the upper panel: 
the experimental data indicate only an approximate scaling at the LHC energy for deuterons.  
To quantify the deviation, the \pt/n$_{\rm q}$ dependence of \vtwo/n$_{\rm q}$ for protons and anti-protons is fitted with a 
seventh-order-polynomial function and the ratio of (\vtwo/n$_{\rm q}$)/(\vtwo/n$_{\rm q}$)$_{\mathrm{Fit\ p}}$ 
is calculated for each particle. A deviation from the n$_{\rm q}$ scaling of the order of 20\% 
for \pt/n$_{\rm q}~>~0.6$ is observed for deuterons; 
the same behaviour is observed in the other centrality intervals (not shown).  Finally, in the third column, the measured \vtwo/n$_{\rm q}$ 
is shown as a function of the transverse kinetic energy scaled by the number of constituent quarks  
($KE_\mathrm{T})/\mathrm{n}_{\rm q} = (m_{\mathrm{T}} - m_0)/\mathrm{n}_{\rm q}$ of each particle.  
This scaling, introduced by the PHENIX collaboration~\cite{Adler:2003kt}  for low \pt, was initially 
observed to work well -- within statistical uncertainties -- at RHIC energies in central A--A collisions \cite{Adams:2003am, Afanasiev:2007tv}. 
However, recent publications report deviations from this scaling for non central Au-Au collisions~\cite{Adare:2012vq}. Also at the LHC
energy the proposed scaling does not work properly (deviations up to~$\sim$~20\%)  \cite{Abelev:2014pua}, and the scaling is not valid for deuterons either. 
The deviations are quantified in the bottom panel, where the ratio of  (\vtwo/n$_{\rm q}$)/(\vtwo/n$_{\rm q}$)$_{\mathrm{Fit\ p}}$ for each particle 
is shown. 
Significant deviations are found for $KE_\mathrm{T}/\mathrm{n}_{\rm q} <$ 0.3 \gmom, indicating that also for light nuclei the scaling with n$_{\rm q}$ does not hold at the LHC energy. For $KE_\mathrm{T}/\mathrm{n}_{\rm q} >$ 0.3 \gmom, data exhibit deviations from an exact scaling at the level of 20\%.  

%% file: comparison_bw.tex
\section{Comparison with different theoretical models}
\label{section:comparisons}
\subsection{Comparison with Blast-Wave model}
The nuclear fireball model was introduced in 1976 to explain 
midrapidity proton-inclusive spectra~\cite{Westfall:1976fu}.  
This model assumes that a clean cylindrical cut is made by the projectile and target leaving a hot 
source in between them. Protons emitted from this fireball  should follow a thermal energy distribution, and are expected to be emitted isotropically.  
Such a model, called Blast-Wave model, has evolved since then, with more parameters to describe both 
the \pt\ spectra and the anisotropic flow of produced particles~\cite{Schnedermann:1993ws, Huovinen:2001cy, Adler:2001nb}.
As described in \cite{Huovinen:2001cy}, the transverse mass spectrum can be expressed as 
\begin{equation}
\frac{\dd\mathrm{N}}{\dd y \dd m_{\rm{T}}^2 \dd\phi_p} \sim 
     \int_0^{2\pi} \!\!\! \dd\phi_s\, 
     K_1(\beta_t(\phi_s))\, e^{\alpha_t(\phi_s)\cos(\phi_s{-}\phi_p)} \,,
\label{eq:spectrum}
\end{equation}
where $\phi_s$  and $\phi_p$ are the azimuthal angles in coordinate and momentum 
space; the arguments $\alpha_t(\phi_s){=}$ $(p_{\mathrm{T}}/T)\sinh(\rho(\phi_s))$  and $\beta_t(\phi_s){=}(m_{\mathrm{T}}/T)\cosh(\rho(\phi_s))$ are based on a 
$\phi_s$-dependent radial flow rapidity $\rho(\phi_s)$ and $K_1$ is a modified Bessel function of the second kind. \\
The elliptic flow coefficient \vtwo\ is obtained by taking the azimuthal average 
over $\cos(2\phi_p)$ with this spectrum, $v_2 = \langle \cos(2\phi_p) \rangle$. The integral on  
$\phi_p$ can be evaluated analytically 
\begin{equation}
 v_2(p_\mathrm{T}) =  \frac{\int_0^{2\pi} \dd\phi_s\ \cos(2\phi_s)\ I_2(\alpha_t(\phi_s)) \ K_1(\beta_t(\phi_s))} 
   {  \int_0^{2\pi} \dd\phi_s\ I_0(\alpha_t(\phi_s))\ K_1(\beta_t(\phi_s))},
\label{eq:v2shell}
\end{equation}
where $I_0$ and $I_2$ are modified Bessel functions of the first kind. 
However, the Blast-Wave fit matched data even better after the STAR Collaboration added a fourth parameter, $s_2$, \cite{Adler:2001nb} which takes
into account the anisotropic shape of the source in coordinate space.  With the introduction of the $s_2$ parameter, the elliptic flow can be expressed as 
\begin{equation}
\label{eq:v2}
v_{2}(p_\mathrm{T}) = \frac{\int_{0}^{2\pi}\dd\phi_{s} \cos(2\phi_{s})I_{2}[\alpha_{t}(\phi_{s})]K_{1}[\beta_{t}(\phi_{s})][1+2s_{2}\cos(2\phi_{s})]}{\int_{0}^{2\pi}\dd\phi_{s} I_{0}[\alpha_{t}(\phi_{s})] K_{1}[\beta_{t}(\phi_{s})][1+2s_{2}\cos(2\phi_{s})]},
\end{equation}
where the masses for different particle species only enter via $m_{T}$ in $\beta_{t}(\phi_{s})$.
The measured pions, kaons and protons \pt\ spectra \cite{Abelev:2013vea} and \vtwo\ (\pt) \cite{Abelev:2014pua} 
are fitted simultaneously using the masses of the different particle species as fixed parameters. 
The parameters extracted from the fit were used to predict deuteron \vtwo(\pt) and \pt\ spectra and are shown in Table~\ref{tab:bwparameters}. The four parameters, as described in   \cite{Adler:2001nb}, represent the kinetic freeze-out temperature ($T$), the mean transverse
expansion rapidity ($\rho_0$), the amplitude of its azimuthal variation ($\rho_a$) and the variation the azimuthal density of the source ($s_2$), respectively.

\begin{table}[!h]
\centering
\begin{tabular}{@{}lccc@{}}
\hline
\hline
\multirow{2}{*}{Fit parameters}        & \multicolumn{3}{c}{Centrality classes} \\
                             & 0-10\%    & 10-20\%       & 20-40\%    \\ 
\hline
$T$ (MeV)                              & $96\pm 3      $  & $ 97\pm 2$       &  $100\pm 2$          \\
$s_2 \times 10^{-2}$                   & $3.21 \pm 0.08$  & $ 6.18 \pm 0.11$   &  $8.97\pm 0.17$   \\
$\rho_0 \times 10^{-1}$    & $8.2 \pm 0.12$   &  $8.18 \pm 0.10$    &  $7.99 \pm 0.12$  \\
$\rho_a \times 10^{-2}$                & $1.21 \pm 0.05$  &  $ 2.25 \pm 0.08$   &  $3.09 \pm 0.11$ \\
\hline
\end{tabular}
\caption{Blast-Wave parameters computed by fitting the pion, kaon and proton transverse-momentum spectra and elliptic flow. See the text for more details.}
\label{tab:bwparameters}
\end{table}

\begin{figure}[!htb]
\begin{tabular}{c}
\begin{minipage}{1.\textwidth}
\centerline{\includegraphics[width=1\textwidth]{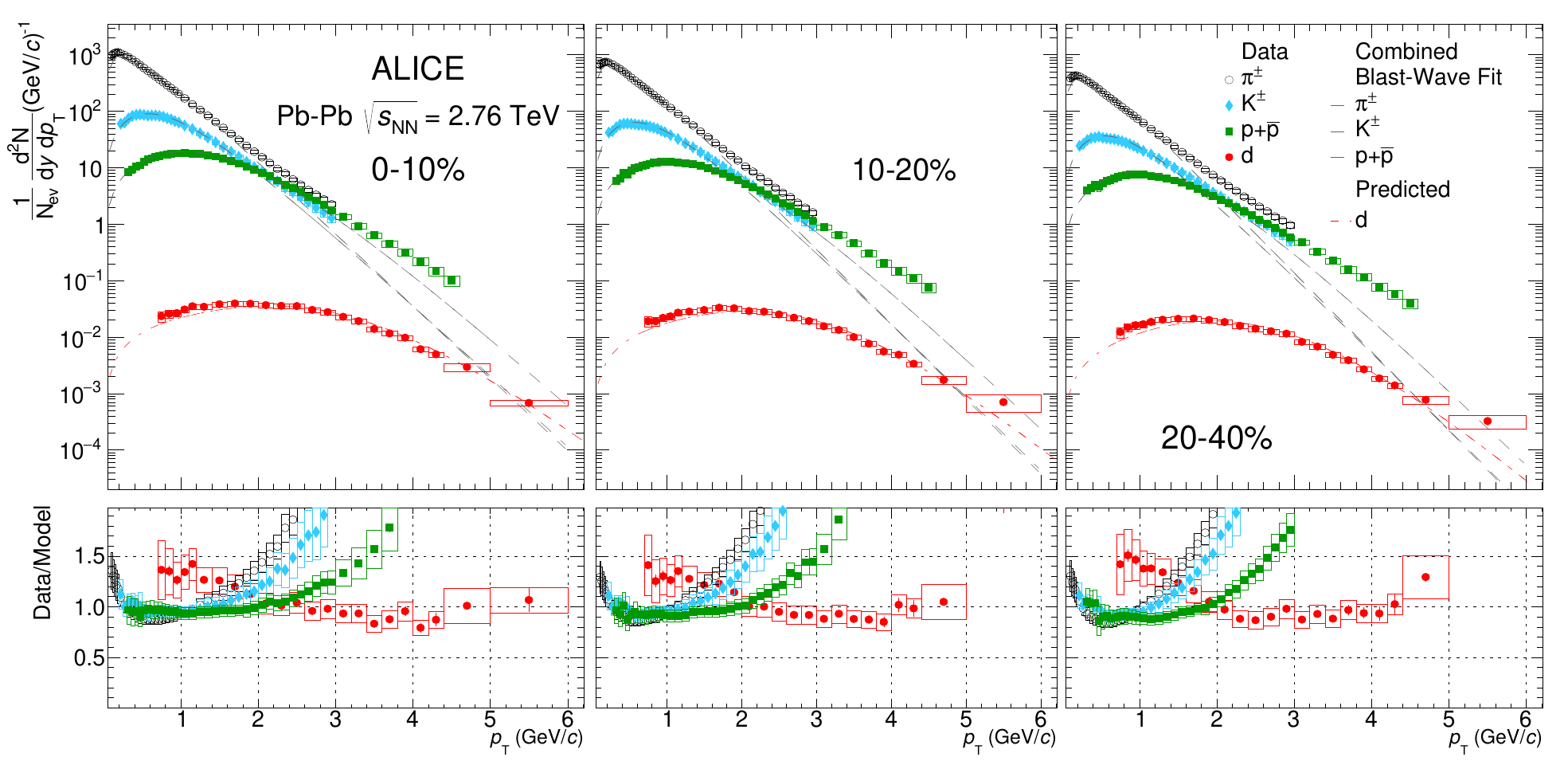}}
\end{minipage}\\

\begin{minipage}{1.\textwidth}
\centerline{\includegraphics[width=1\textwidth]{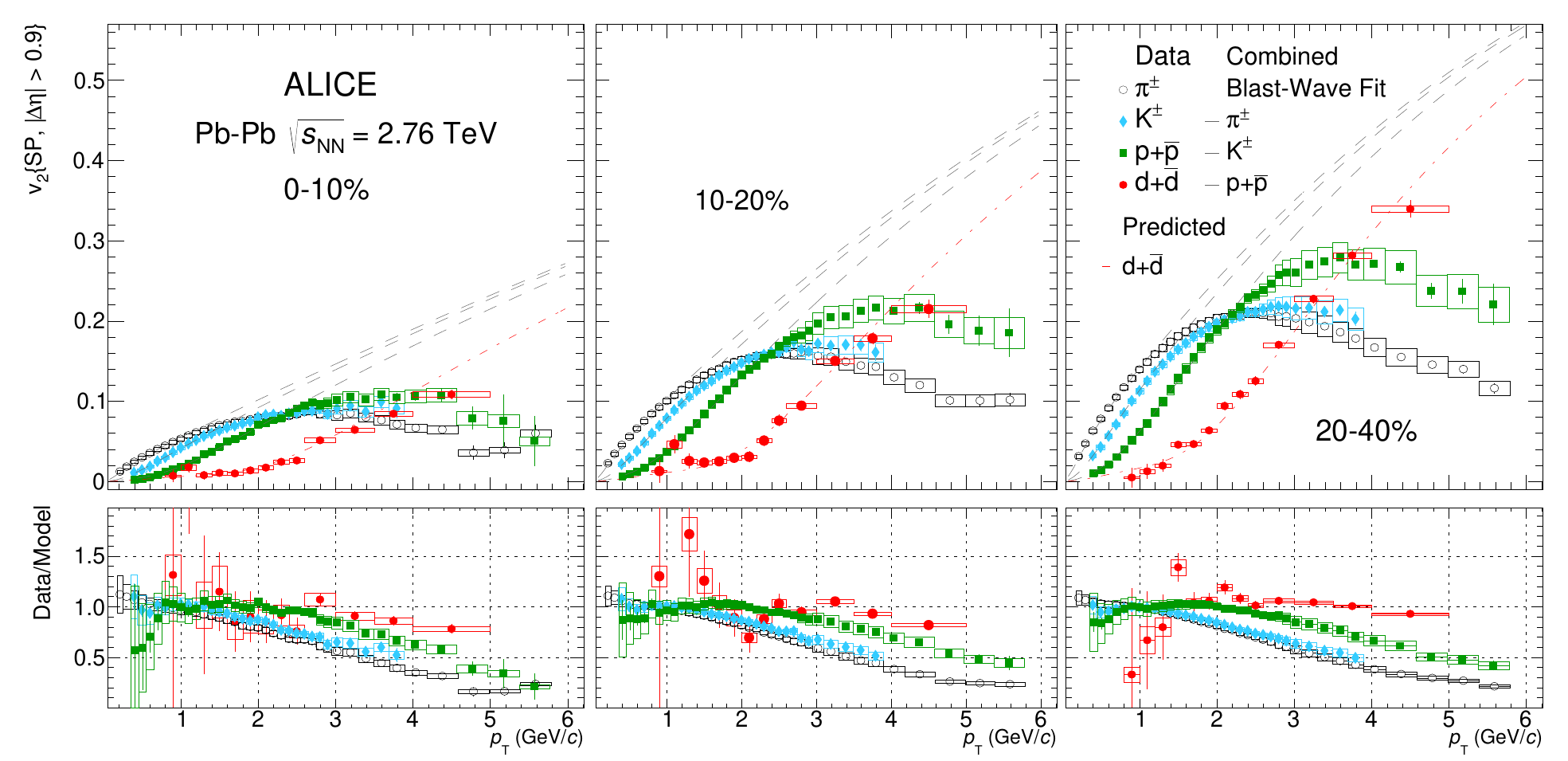}}
\end{minipage} \\

\end{tabular}
\caption{Combined Blast-Wave fit to \pt\ and \vtwo(\pt) distributions using 
equations \ref{eq:spectrum} and \ref{eq:v2}. The six upper 
panels show the \pt\ spectra and the ratio between data and fit, while the 
six panels in the bottom part shows the \vtwo(\pt) and the ratio between 
data and fit. In each panel, $\pi^\pm$ (empty circles), K$^\pm$ (diamonds), 
p+$\overline{\rm p}$ (filled squares) and d+\dbar\ (filled circles) are shown. For 
$\pi^\pm$, K$^\pm$ and p+$\overline{\rm p}$ the long dashed 
curves represent the combined \pt\ and \vtwo\ Blast-Wave fit. Deuteron curves (dash dotted lines) 
are predictions from lighter particles Blast-Wave combined fit. Each column shows a different centrality 
intervals (0-10\% left, 10-20\% middle and 20-40\% right). }
\label{fig:BWComparison}
\end{figure}

The simultaneous fit to \pt\ spectra and \vtwo\ (\pt) and the predictions for 
deuterons are shown in Figure~\ref{fig:BWComparison}; the centrality decreases going from the left to the right. 
In the upper part of Figure~\ref{fig:BWComparison} the \pt\ spectra, as well as the  ratio between 
data and model for different centrality intervals, are shown, while the bottom part of the 
Figure~\ref{fig:BWComparison} shows the \vtwo(\pt) and the ratio between data and model for several 
centrality intervals. 
The transverse momentum intervals where the different particle species were fitted are 
\mbox{[0.5-1]~\gmom} for pions, \mbox{[0.2-1.2]~\gmom} for kaons and \mbox{[0.3-1.7]~\gmom} for protons. 
These ranges were chosen to be similar to what shown in ~\cite{Adam:2015vda} and to be able to fit at 
the same time transverse-momentum spectra and \vtwo\ distributions.
As can be observed in Figure~\ref{fig:BWComparison}, the combined fit gives a good description  of the 
deuterons \vtwo(\pt) within the statistical uncertainties for all measured transverse momenta and centralities. This is in contrast to what has been observed by the STAR experiment in Au--Au collisions at \s$=200$ GeV \cite{Adamczyk:2016gfs}, where the Blast-Wave  model underestimates the deuteron \vtwo\ measured in data.
Deuteron spectra are underestimated at low \pt\ (deviations up to 2$\sigma$ for \pt\ smaller than 1.8~\gmom), 
while the model is able to reproduce the measured data within 1 $\sigma$ for \pt\ up to 6~\gmom.

%% file: comparison_coalescence.tex
\subsection{Comparison with coalescence model}

\begin{figure}[!htbp]
\begin{tabular}{ccc}
\begin{minipage}{.5\textwidth}
\centerline{\includegraphics[width=1\textwidth]{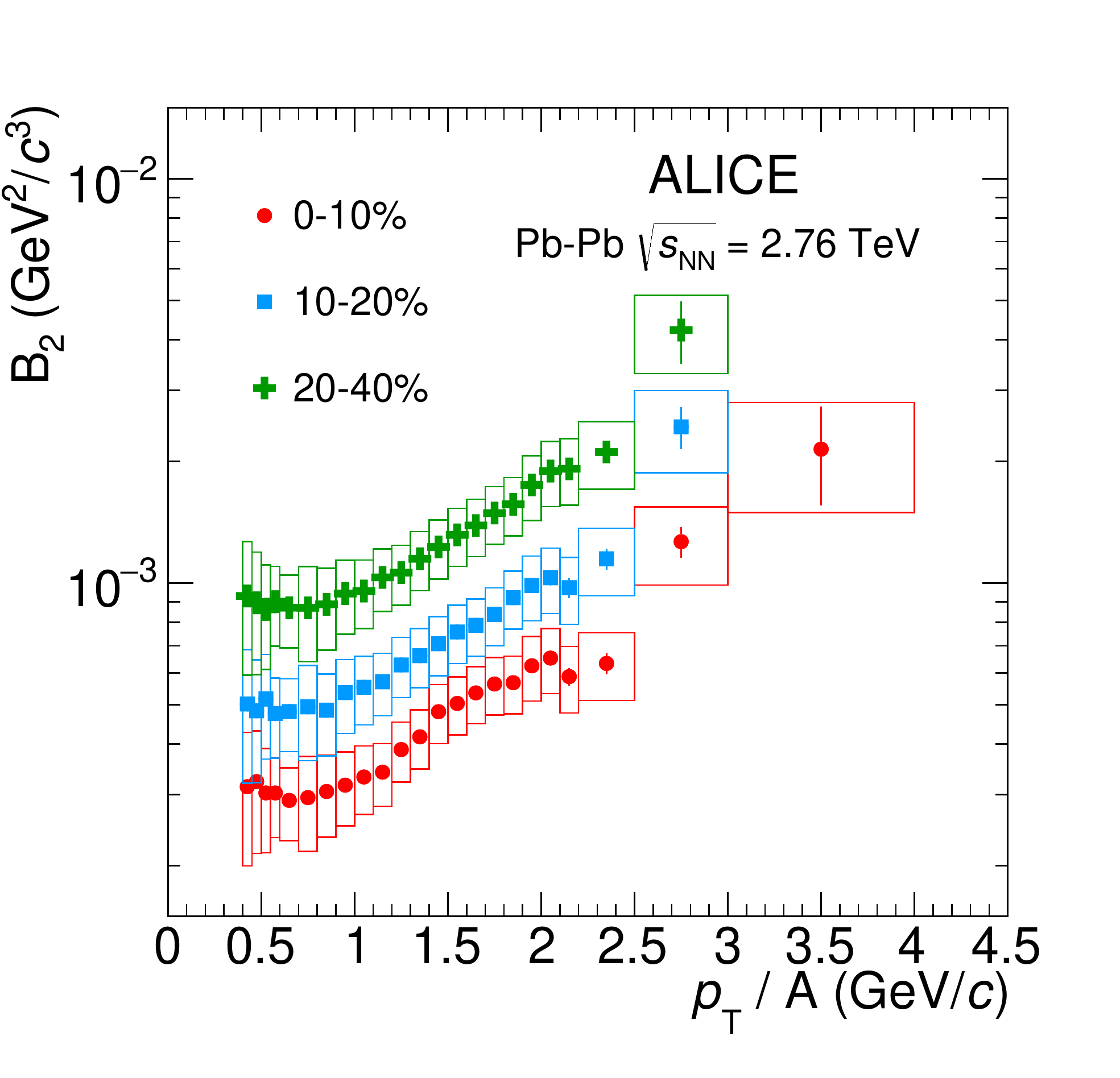}}
\end{minipage} & 
\begin{minipage}{.5\textwidth}
\centerline{\includegraphics[width=1\textwidth]{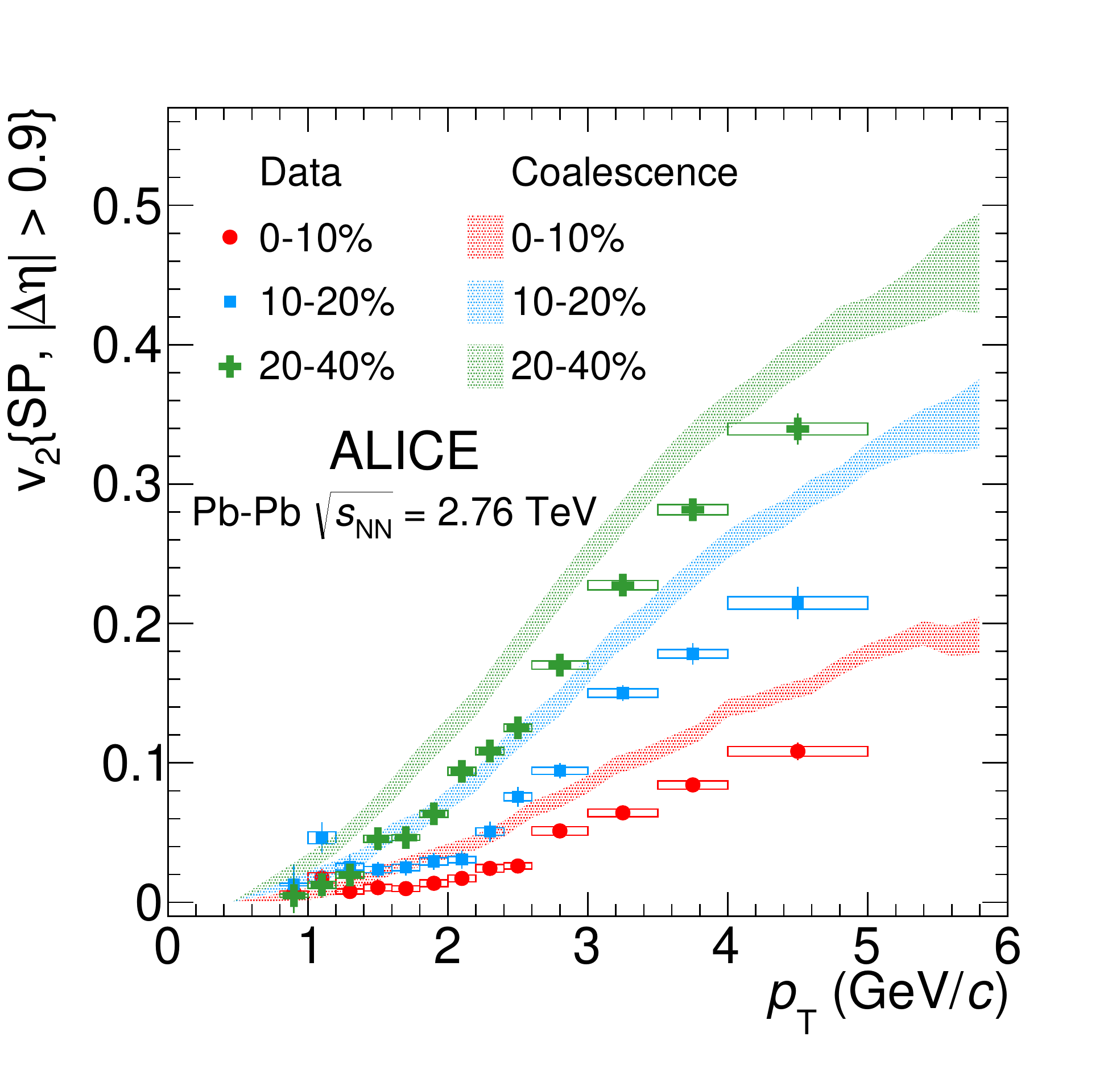}}
\end{minipage} 
\end{tabular}
\caption{Left: Coalescence parameter $B_2$ as a function of the transverse momentum per nucleon (\pt/A) for 
different centrality classes. Right: Measured \vtwo\ of d+\dbar\ compared with the expectations from simple 
coalescence (equation~\ref{eq:v2coal}) for different centrality intervals as indicated in the legend.}
\label{fig:CoalesceDeuteron}
\end{figure}

Light nuclei have nucleons as constituents and it has been supposed that they are 
likely to be formed via coalescence of protons and neutrons which are close in space and have similar velocities. 
In this production mechanism, the cross section for the production 
of a cluster with mass number $A$ is related to the probability that $A$ nucleons have relative 
momenta less than $p_0$, which is a free parameter of the model \cite{Csernai:1986qf}.  
This provides the following relation between the production rate of the nuclear 
cluster emitted with a momentum $p_A$ and the nucleons emitted with a momentum $p_{\mathrm{p}}$

\begin{equation}
E_A \frac{\mathrm{d}^3N_A}{\mathrm{d}p_A^3} = B_A \left(E_{\mathrm{p}} \frac{\mathrm{d}^3N_{\mathrm{p}}}{\mathrm{d}p_{\mathrm{p}}^3}\right)^A,
\label{eq:coal}
\end{equation}
where $p_A = Ap_{\mathrm{p}}$. For a given nucleus, if the spin factors are neglected, 
the coalescence parameter $B_A$ does not depend on the momentum since it depends only 
on the cluster parameters 
\begin{equation}
B_A =  \left( \frac{4 \pi}{3} p_0^3 \right) ^{(A-1)} \frac{1}{A!} \frac{M}{m^A},
\label{eq:coalescece}
\end{equation}
where $p_0$ is commonly named coalescence radius while $M$ and $m$ are the nucleus and the nucleon mass, respectively.
The left panel of Figure~\ref{fig:CoalesceDeuteron} shows the obtained 
$B_2$ values for deuterons in three different centrality regions 
studied in the present work. The measured $B_2$ values are plotted versus the transverse momentum per
nucleon (\pt/$A$). A clear decrease of the $B_2$ parameter with increasing centrality and an increase with 
transverse momentum is observed. The measured $B_2$ at higher \pt\ bins presented in this paper follow 
the trend already observed for smaller momenta, confirming that the experimental result is 
in contrast to the expectations of the simplest coalescence model~\cite{Csernai:1986qf}, 
where the $B_2$ is expected to be flat. As already observed in \cite{Adam:2015vda}, the observed behaviour
can be qualitatively explained by position-momentum correlations which are caused by a radially expanding 
source \cite{Polleri:1997bp}, but better theoretical calculations at the LHC energies are needed.\\ 
Since elliptic flow is additive, it is possible to infer the expected \vtwo\ of a composite 
state (like a deuteron) formed via coalescence starting from equation~\ref{eq:coal}. 
In the region where the coalescence occurs, the elliptic flow of a nucleus can be expressed as a function of the 
the elliptic flow of its constituent nucleons. For a deuteron, assuming that protons and neutrons behave in the same way, the following relation is expected~\cite{Molnar:2003ff}:
\begin{equation}
v_{2,\mathrm{d}}(p_{\rm T}) = \frac{2 v_{2,\mathrm{p}} (p_{\rm {T}}/2)}{1+2{v^2_{2,\rm{p}}} (p_{\mathrm {T}}/2)}.
\label{eq:v2coal}
\end{equation} 
It is then possible to obtain the expected deuteron elliptic flow starting from the one
measured for protons~\cite{Abelev:2014pua}. The results for different centrality intervals 
are shown in the right part of Figure~\ref{fig:CoalesceDeuteron}, where the measured
elliptic flow (markers) is compared with simple coalescence predictions (shaded bands) from 
equation~\ref{eq:v2coal} for the three different centrality intervals presented in the paper. 
Also here the simple coalescence is not able to reproduce the measured elliptic
flow of deuterons. This behaviour is different with respect to what has been observed at lower energies, where an atomic mass number scaling was observed in the 0.3~$<$~\pt~$<$~3~\gmom\ interval~\cite{Adamczyk:2016gfs}. 
Improved versions of the coalescence model, for instance based on a realistic phase space distribution of the constituent protons and neutrons, might describe the elliptic flow of deuterons at LHC energy better.

%% file: comparison_ampt.tex
\subsection{Comparison with AMPT}
\begin{figure}[!htb]
\begin{tabular}{ccc}
\begin{minipage}{.5\textwidth}
\centerline{\includegraphics[width=1\textwidth]{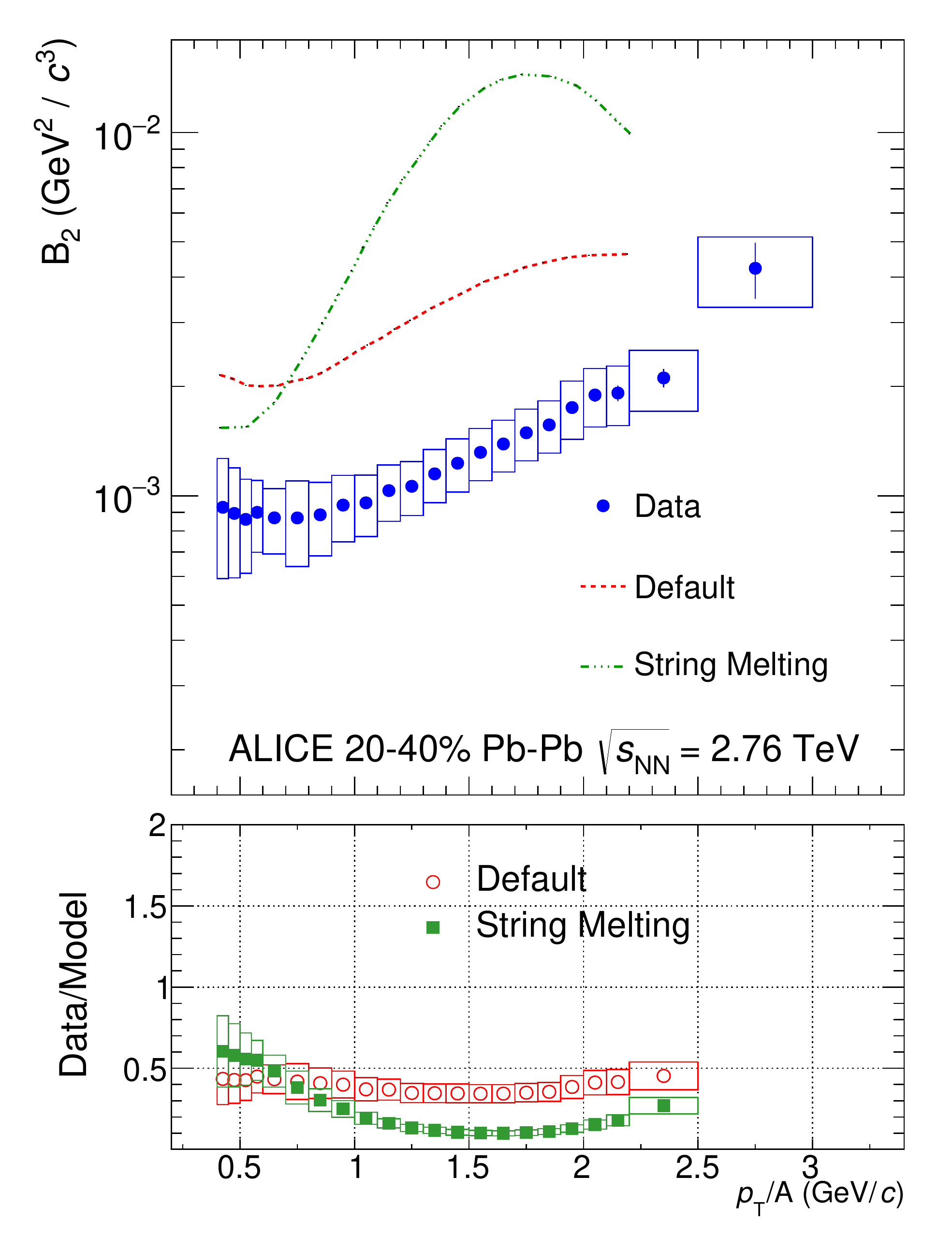}}
\end{minipage} & 
\begin{minipage}{.5\textwidth}
\centerline{\includegraphics[width=1\textwidth]{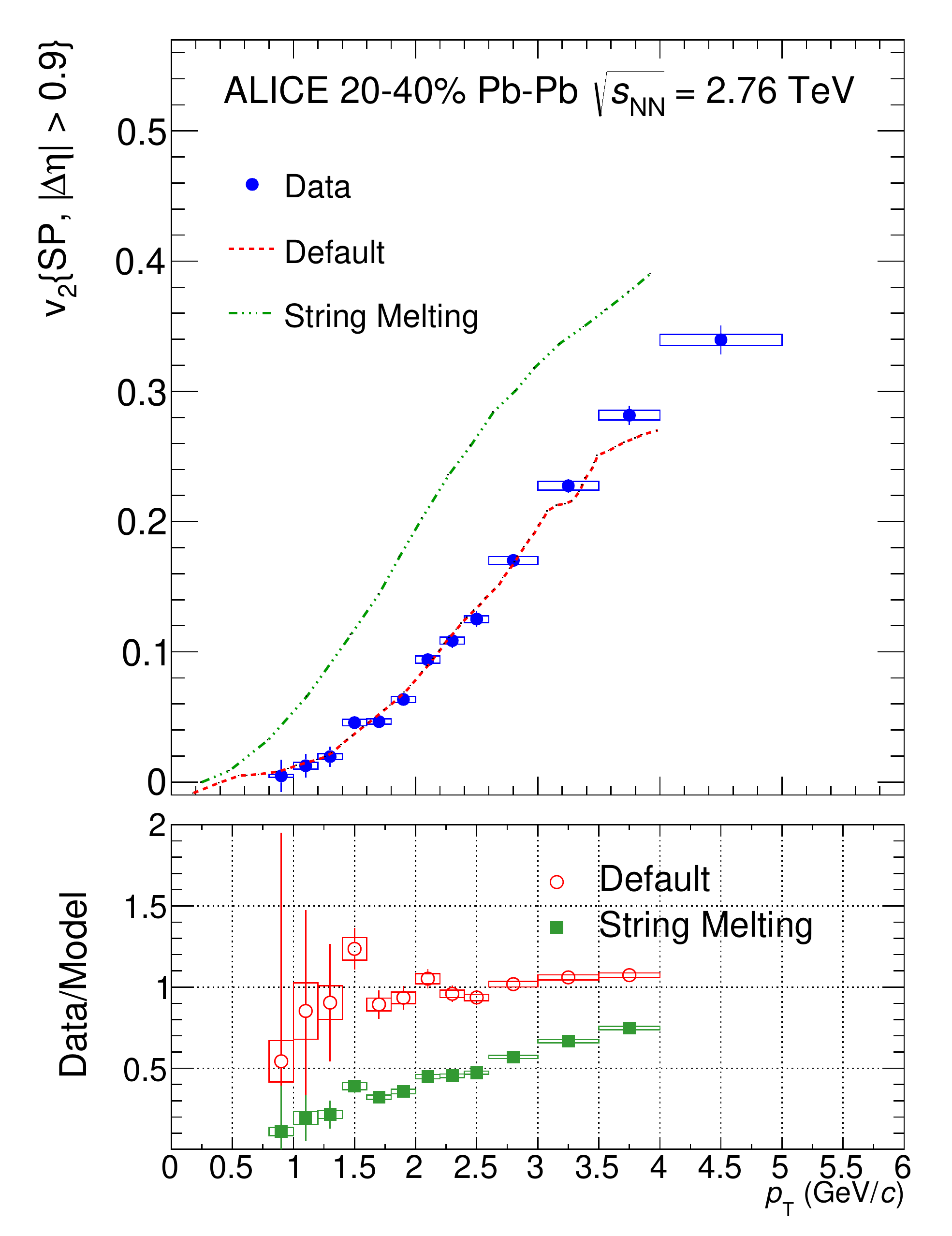}}
\end{minipage} 
\end{tabular}
\caption{Measured deuteron (filled dots) $B_2$ as a function of the transverse momentum 
per nucleon (\pt/A) (top left panel) and \vtwo(\pt) parameter (top right panel)  
compared to those produced with two versions of the AMPT model (dashed and dotted lines). 
In the bottom panels, the ratios between the measured data and the expectations from 
the two models are shown.}
\label{fig:ComparisonAMPT}
\end{figure}

The AMPT model is a hybrid model \cite{Lin:2004en} with the initial particle distributions generated 
with HIJING~\cite{Wang:1991h}.  
In the default version of AMPT, the jet quenching in HIJING is replaced by explicitly taking into 
account the scattering of mini-jet partons via the Zhang's parton cascade (ZPC) model \cite{Zhang:1997ej}.
In the version with string melting, all the hadrons produced from the 
string fragmentation in HIJING are converted to their
valence quarks and antiquarks, whose evolution in time
and space is guided by the ZPC model. 
After the end of their scatterings, quarks and antiquarks are converted to
hadrons via a spatial coalescence model. 
In both versions of the AMPT model, the scatterings among hadrons are
described by a relativistic transport (ART) model \cite{Li:1995pra}.
The (anti-)deuterons are produced and dissolved within ART via the  
$\mathrm{NN}  \leftrightarrow \pi \mathrm{d}$  reaction in the hadronic stage of AMPT.
The Monte Carlo predictions~\cite{Zhu:2015voa} for the deuteron coalescence parameter 
($B_2$) and elliptic flow (\vtwo) are compared with the measured one in this section.
For the simulation, an impact parameter $b$ = 8 fm was used: this value corresponds to 
the mean value of the impact parameter in the 20--40\% centrality interval \cite{Abelev:2013qoq}.  
The comparisons between data and Monte Carlo results 
can be seen in Figure~\ref{fig:ComparisonAMPT}: in the left top panel  
the measured $B_2$ is shown (filled markers) together with the published AMPT expectations 
(lines), while in the bottom panel the ratios between the data and models are shown. 
In the top right panel the measured elliptic flow (filled markers) and the AMPT 
expectations are displayed.  
From the bottom right panel of Figure~\ref{fig:ComparisonAMPT} it is possible to 
observe that the default version of AMPT (empty circles) is able to reproduce the measured 
deuteron elliptic flow, while the simulated $B_2$ is able to reproduce the behaviour of the measured 
one but is larger by a factor 2.  
The AMPT with String Melting enabled (dotted line in the top panels and full squares in
the bottom panels) is unable to reproduce neither the measured $B_2$ nor the \vtwo(\pt) parameter
of the (anti-)deuterons.
It is worth to emphasize that the standart AMPT does not reproduce the lighter hadrons \vtwo(\pt) neither~\cite{Abelev:2014pua}.

%% file: conclusion.tex
\section{Conclusions}
\label{sec:conclusions}
In this paper we presented the deuteron spectra up to $\pt~=~8$~\gmom\ 
produced in \PbPb\ collisions at \s~=~2.76~TeV  
extending significantly the \pt\ reached by the spectra shown in
\cite{Adam:2015vda} for central and semi-central collisions. 
The \vtwo\ of (anti-)deuterons is measured up to 5 \gmom\ and it shows an increasing trend going from central 
to semi-central collisions, consistent with the picture of the final state anisotropy driven by the collision geometry. 
At low \pt\ the deuteron \vtwo\ follows the mass ordering, indicating a more pronounced radial flow in the most 
central collisions, as it is observed also for lighter particles. Similarly the n$_{\mathrm{q}}$ scaling violation seen for 
the elliptic flow of lighter particles at the LHC energies is confirmed for the elliptic flow of the deuteron.
The Blast-Wave model, fitted on the spectra and the elliptic flow of pions, kaons and protons, describes within 
the experimental uncertainties the deuteron production spectra for \pt $>$ 1.8 \gmom ~and the deuteron  \vtwo(\pt) 
suggesting common kinetic freeze-out conditions. At lower transverse momenta the model underestimates the measured spectra with a discrepancy up to $2\sigma$.\\
The coalescence parameter $B_2$ , evaluated up to \pt$/A$ $= 4$ \gmom, rapidly increases with the transverse 
momentum confirming the experimental observation made in \cite{Adam:2015vda}. The simplest formulation of 
the coalescence model ~\cite{Csernai:1986qf} predicts a flat $B_2$ distribution and it does not reproduce the observed trend. 
The same model fails to reproduce the measured elliptic flow of deuterons.
On the contrary, the AMPT model without String Melting \cite{Lin:2004en} is able to reproduce the observed 
elliptic flow of deuterons and it correctly predicts the shape of the $B_2$ distribution but it overestimates the data 
by about a factor of 2. When enabling the String Melting mechanism the AMPT model is unable to predict 
neither the measured $B_2$ nor the \vtwo(\pt) of the (anti-)deuterons. 

%% file: fa_2017-06-12.tex

The ALICE Collaboration would like to thank all its engineers and technicians for their invaluable contributions to the construction of the experiment and the CERN accelerator teams for the outstanding performance of the LHC complex.
The ALICE Collaboration gratefully acknowledges the resources and support provided by all Grid centres and the Worldwide LHC Computing Grid (WLCG) collaboration.
The ALICE Collaboration acknowledges the following funding agencies for their support in building and running the ALICE detector:
A. I. Alikhanyan National Science Laboratory (Yerevan Physics Institute) Foundation (ANSL), State Committee of Science and World Federation of Scientists (WFS), Armenia;
Austrian Academy of Sciences and Nationalstiftung f\"{u}r Forschung, Technologie und Entwicklung, Austria;
Ministry of Communications and High Technologies, National Nuclear Research Center, Azerbaijan;
Conselho Nacional de Desenvolvimento Cient\'{\i}fico e Tecnol\'{o}gico (CNPq), Universidade Federal do Rio Grande do Sul (UFRGS), Financiadora de Estudos e Projetos (Finep) and Funda\c{c}\~{a}o de Amparo \`{a} Pesquisa do Estado de S\~{a}o Paulo (FAPESP), Brazil;
Ministry of Science \& Technology of China (MSTC), National Natural Science Foundation of China (NSFC) and Ministry of Education of China (MOEC) , China;
Ministry of Science, Education and Sport and Croatian Science Foundation, Croatia;
Ministry of Education, Youth and Sports of the Czech Republic, Czech Republic;
The Danish Council for Independent Research | Natural Sciences, the Carlsberg Foundation and Danish National Research Foundation (DNRF), Denmark;
Helsinki Institute of Physics (HIP), Finland;
Commissariat \`{a} l'Energie Atomique (CEA) and Institut National de Physique Nucl\'{e}aire et de Physique des Particules (IN2P3) and Centre National de la Recherche Scientifique (CNRS), France;
Bundesministerium f\"{u}r Bildung, Wissenschaft, Forschung und Technologie (BMBF) and GSI Helmholtzzentrum f\"{u}r Schwerionenforschung GmbH, Germany;
General Secretariat for Research and Technology, Ministry of Education, Research and Religions, Greece;
National Research, Development and Innovation Office, Hungary;
Department of Atomic Energy Government of India (DAE) and Council of Scientific and Industrial Research (CSIR), New Delhi, India;
Indonesian Institute of Science, Indonesia;
Centro Fermi - Museo Storico della Fisica e Centro Studi e Ricerche Enrico Fermi and Istituto Nazionale di Fisica Nucleare (INFN), Italy;
Institute for Innovative Science and Technology , Nagasaki Institute of Applied Science (IIST), Japan Society for the Promotion of Science (JSPS) KAKENHI and Japanese Ministry of Education, Culture, Sports, Science and Technology (MEXT), Japan;
Consejo Nacional de Ciencia (CONACYT) y Tecnolog\'{i}a, through Fondo de Cooperaci\'{o}n Internacional en Ciencia y Tecnolog\'{i}a (FONCICYT) and Direcci\'{o}n General de Asuntos del Personal Academico (DGAPA), Mexico;
Nederlandse Organisatie voor Wetenschappelijk Onderzoek (NWO), Netherlands;
The Research Council of Norway, Norway;
Commission on Science and Technology for Sustainable Development in the South (COMSATS), Pakistan;
Pontificia Universidad Cat\'{o}lica del Per\'{u}, Peru;
Ministry of Science and Higher Education and National Science Centre, Poland;
Korea Institute of Science and Technology Information and National Research Foundation of Korea (NRF), Republic of Korea;
Ministry of Education and Scientific Research, Institute of Atomic Physics and Romanian National Agency for Science, Technology and Innovation, Romania;
Joint Institute for Nuclear Research (JINR), Ministry of Education and Science of the Russian Federation and National Research Centre Kurchatov Institute, Russia;
Ministry of Education, Science, Research and Sport of the Slovak Republic, Slovakia;
National Research Foundation of South Africa, South Africa;
Centro de Aplicaciones Tecnol\'{o}gicas y Desarrollo Nuclear (CEADEN), Cubaenerg\'{\i}a, Cuba, Ministerio de Ciencia e Innovacion and Centro de Investigaciones Energ\'{e}ticas, Medioambientales y Tecnol\'{o}gicas (CIEMAT), Spain;
Swedish Research Council (VR) and Knut \& Alice Wallenberg Foundation (KAW), Sweden;
European Organization for Nuclear Research, Switzerland;
National Science and Technology Development Agency (NSDTA), Suranaree University of Technology (SUT) and Office of the Higher Education Commission under NRU project of Thailand, Thailand;
Turkish Atomic Energy Agency (TAEK), Turkey;
National Academy of  Sciences of Ukraine, Ukraine;
Science and Technology Facilities Council (STFC), United Kingdom;
National Science Foundation of the United States of America (NSF) and United States Department of Energy, Office of Nuclear Physics (DOE NP), United States of America.

%% file: Alice_Authorlist_2017-Jun-12.tex


\begingroup
\small
\begin{flushleft}
S.~Acharya$^\textrm{\scriptsize 139}$,
D.~Adamov\'{a}$^\textrm{\scriptsize 96}$,
J.~Adolfsson$^\textrm{\scriptsize 34}$,
M.M.~Aggarwal$^\textrm{\scriptsize 101}$,
G.~Aglieri Rinella$^\textrm{\scriptsize 35}$,
M.~Agnello$^\textrm{\scriptsize 31}$,
N.~Agrawal$^\textrm{\scriptsize 48}$,
Z.~Ahammed$^\textrm{\scriptsize 139}$,
N.~Ahmad$^\textrm{\scriptsize 17}$,
S.U.~Ahn$^\textrm{\scriptsize 80}$
% ...
\renewcommand\labelenumi{\textsuperscript{\theenumi}~}

\section*{Affiliation notes}
\renewcommand\theenumi{\roman{enumi}}
\begin{Authlist}
\item \Adef{0}Deceased
\item \Adef{idp1786352}{Also at: Dipartimento DET del Politecnico di Torino, Turin, Italy}
\item \Adef{idp1805744}{Also at: Georgia State University, Atlanta, Georgia, United States}
\item \Adef{idp4096400}{Also at: M.V. Lomonosov Moscow State University, D.V. Skobeltsyn Institute of Nuclear, Physics, Moscow, Russia}
\item \Adef{idp4453824}{Also at: Department of Applied Physics, Aligarh Muslim University, Aligarh, India}
\item \Adef{idp5429952}{Also at: Institute of Theoretical Physics, University of Wroclaw, Poland}
\end{Authlist}

\section*{Collaboration Institutes}
\renewcommand\theenumi{\arabic{enumi}~}

$^{1}$A.I. Alikhanyan National Science Laboratory (Yerevan Physics Institute) Foundation, Yerevan, Armenia
\\
$^{2}$Benem\'{e}rita Universidad Aut\'{o}noma de Puebla, Puebla, Mexico
\\
$^{3}$Bogolyubov Institute for Theoretical Physics, Kiev, Ukraine
\\
$^{4}$Bose Institute, Department of Physics 
and Centre for Astroparticle Physics and Space Science (CAPSS), Kolkata, India
\\
$^{5}$Budker Institute for Nuclear Physics, Novosibirsk, Russia
\\
$^{6}$California Polytechnic State University, San Luis Obispo, California, United States
\\
$^{7}$Central China Normal University, Wuhan, China
\\
$^{8}$Centre de Calcul de l'IN2P3, Villeurbanne, Lyon, France
\\
$^{9}$Centro de Aplicaciones Tecnol\'{o}gicas y Desarrollo Nuclear (CEADEN), Havana, Cuba
\\
$^{10}$Centro de Investigaciones Energ\'{e}ticas Medioambientales y Tecnol\'{o}gicas (CIEMAT), Madrid, Spain
\\
$^{11}$Centro de Investigaci\'{o}n y de Estudios Avanzados (CINVESTAV), Mexico City and M\'{e}rida, Mexico
\\
$^{12}$Centro Fermi - Museo Storico della Fisica e Centro Studi e Ricerche ``Enrico Fermi', Rome, Italy
\\
$^{13}$Chicago State University, Chicago, Illinois, United States
\\
$^{14}$China Institute of Atomic Energy, Beijing, China
\\
$^{15}$COMSATS Institute of Information Technology (CIIT), Islamabad, Pakistan
\\
$^{16}$Departamento de F\'{\i}sica de Part\'{\i}culas and IGFAE, Universidad de Santiago de Compostela, Santiago de Compostela, Spain
\\
$^{17}$Department of Physics, Aligarh Muslim University, Aligarh, India
\\
$^{18}$Department of Physics, Ohio State University, Columbus, Ohio, United States
\\
$^{19}$Department of Physics, Pusan National University, Pusan, Republic of Korea
\\
$^{20}$Department of Physics, Sejong University, Seoul, Republic of Korea
\\
$^{21}$Department of Physics, University of Oslo, Oslo, Norway
\\
$^{22}$Department of Physics and Technology, University of Bergen, Bergen, Norway
\\
$^{23}$Dipartimento di Fisica dell'Universit\`{a} 'La Sapienza'
and Sezione INFN, Rome, Italy
\\
$^{24}$Dipartimento di Fisica dell'Universit\`{a}
and Sezione INFN, Cagliari, Italy
\\
$^{25}$Dipartimento di Fisica dell'Universit\`{a}
and Sezione INFN, Trieste, Italy
\\
$^{26}$Dipartimento di Fisica dell'Universit\`{a}
and Sezione INFN, Turin, Italy
\\
$^{27}$Dipartimento di Fisica e Astronomia dell'Universit\`{a}
and Sezione INFN, Bologna, Italy
\\
$^{28}$Dipartimento di Fisica e Astronomia dell'Universit\`{a}
and Sezione INFN, Catania, Italy
\\
$^{29}$Dipartimento di Fisica e Astronomia dell'Universit\`{a}
and Sezione INFN, Padova, Italy
\\
$^{30}$Dipartimento di Fisica `E.R.~Caianiello' dell'Universit\`{a}
and Gruppo Collegato INFN, Salerno, Italy
\\
$^{31}$Dipartimento DISAT del Politecnico and Sezione INFN, Turin, Italy
\\
$^{32}$Dipartimento di Scienze e Innovazione Tecnologica dell'Universit\`{a} del Piemonte Orientale and INFN Sezione di Torino, Alessandria, Italy
\\
$^{33}$Dipartimento Interateneo di Fisica `M.~Merlin'
and Sezione INFN, Bari, Italy
\\
$^{34}$Division of Experimental High Energy Physics, University of Lund, Lund, Sweden
\\
$^{35}$European Organization for Nuclear Research (CERN), Geneva, Switzerland
\\
$^{36}$Excellence Cluster Universe, Technische Universit\"{a}t M\"{u}nchen, Munich, Germany
\\
$^{37}$Faculty of Engineering, Bergen University College, Bergen, Norway
\\
$^{38}$Faculty of Mathematics, Physics and Informatics, Comenius University, Bratislava, Slovakia
\\
$^{39}$Faculty of Nuclear Sciences and Physical Engineering, Czech Technical University in Prague, Prague, Czech Republic
\\
$^{40}$Faculty of Science, P.J.~\v{S}af\'{a}rik University, Ko\v{s}ice, Slovakia
\\
$^{41}$Faculty of Technology, Buskerud and Vestfold University College, Tonsberg, Norway
\\
$^{42}$Frankfurt Institute for Advanced Studies, Johann Wolfgang Goethe-Universit\"{a}t Frankfurt, Frankfurt, Germany
\\
$^{43}$Gangneung-Wonju National University, Gangneung, Republic of Korea
\\
$^{44}$Gauhati University, Department of Physics, Guwahati, India
\\
$^{45}$Helmholtz-Institut f\"{u}r Strahlen- und Kernphysik, Rheinische Friedrich-Wilhelms-Universit\"{a}t Bonn, Bonn, Germany
\\
$^{46}$Helsinki Institute of Physics (HIP), Helsinki, Finland
\\
$^{47}$Hiroshima University, Hiroshima, Japan
\\
$^{48}$Indian Institute of Technology Bombay (IIT), Mumbai, India
\\
$^{49}$Indian Institute of Technology Indore, Indore, India
\\
$^{50}$Indonesian Institute of Sciences, Jakarta, Indonesia
\\
$^{51}$INFN, Laboratori Nazionali di Frascati, Frascati, Italy
\\
$^{52}$INFN, Laboratori Nazionali di Legnaro, Legnaro, Italy
\\
$^{53}$INFN, Sezione di Bari, Bari, Italy
\\
$^{54}$INFN, Sezione di Bologna, Bologna, Italy
\\
$^{55}$INFN, Sezione di Cagliari, Cagliari, Italy
\\
$^{56}$INFN, Sezione di Catania, Catania, Italy
\\
$^{57}$INFN, Sezione di Padova, Padova, Italy
\\
$^{58}$INFN, Sezione di Roma, Rome, Italy
\\
$^{59}$INFN, Sezione di Torino, Turin, Italy
\\
$^{60}$INFN, Sezione di Trieste, Trieste, Italy
\\
$^{61}$Inha University, Incheon, Republic of Korea
\\
$^{62}$Institut de Physique Nucl\'eaire d'Orsay (IPNO), Universit\'e Paris-Sud, CNRS-IN2P3, Orsay, France
\\
$^{63}$Institute for Nuclear Research, Academy of Sciences, Moscow, Russia
\\
$^{64}$Institute for Subatomic Physics of Utrecht University, Utrecht, Netherlands
\\
$^{65}$Institute for Theoretical and Experimental Physics, Moscow, Russia
\\
$^{66}$Institute of Experimental Physics, Slovak Academy of Sciences, Ko\v{s}ice, Slovakia
\\
$^{67}$Institute of Physics, Academy of Sciences of the Czech Republic, Prague, Czech Republic
\\
$^{68}$Institute of Physics, Bhubaneswar, India
\\
$^{69}$Institute of Space Science (ISS), Bucharest, Romania
\\
$^{70}$Institut f\"{u}r Informatik, Johann Wolfgang Goethe-Universit\"{a}t Frankfurt, Frankfurt, Germany
\\
$^{71}$Institut f\"{u}r Kernphysik, Johann Wolfgang Goethe-Universit\"{a}t Frankfurt, Frankfurt, Germany
\\
$^{72}$Institut f\"{u}r Kernphysik, Westf\"{a}lische Wilhelms-Universit\"{a}t M\"{u}nster, M\"{u}nster, Germany
\\
$^{73}$Instituto de Ciencias Nucleares, Universidad Nacional Aut\'{o}noma de M\'{e}xico, Mexico City, Mexico
\\
$^{74}$Instituto de F\'{i}sica, Universidade Federal do Rio Grande do Sul (UFRGS), Porto Alegre, Brazil
\\
$^{75}$Instituto de F\'{\i}sica, Universidad Nacional Aut\'{o}noma de M\'{e}xico, Mexico City, Mexico
\\
$^{76}$IRFU, CEA, Universit\'{e} Paris-Saclay, Saclay, France
\\
$^{77}$iThemba LABS, National Research Foundation, Somerset West, South Africa
\\
$^{78}$Joint Institute for Nuclear Research (JINR), Dubna, Russia
\\
$^{79}$Konkuk University, Seoul, Republic of Korea
\\
$^{80}$Korea Institute of Science and Technology Information, Daejeon, Republic of Korea
\\
$^{81}$KTO Karatay University, Konya, Turkey
\\
$^{82}$Laboratoire de Physique Corpusculaire (LPC), Clermont Universit\'{e}, Universit\'{e} Blaise Pascal, CNRS--IN2P3, Clermont-Ferrand, France
\\
$^{83}$Laboratoire de Physique Subatomique et de Cosmologie, Universit\'{e} Grenoble-Alpes, CNRS-IN2P3, Grenoble, France
\\
$^{84}$Lawrence Berkeley National Laboratory, Berkeley, California, United States
\\
$^{85}$Moscow Engineering Physics Institute, Moscow, Russia
\\
$^{86}$Nagasaki Institute of Applied Science, Nagasaki, Japan
\\
$^{87}$National and Kapodistrian University of Athens, Physics Department, Athens, Greece
\\
$^{88}$National Centre for Nuclear Studies, Warsaw, Poland
\\
$^{89}$National Institute for Physics and Nuclear Engineering, Bucharest, Romania
\\
$^{90}$National Institute of Science Education and Research, HBNI, Jatni, India
\\
$^{91}$National Nuclear Research Center, Baku, Azerbaijan
\\
$^{92}$National Research Centre Kurchatov Institute, Moscow, Russia
\\
$^{93}$Niels Bohr Institute, University of Copenhagen, Copenhagen, Denmark
\\
$^{94}$Nikhef, Nationaal instituut voor subatomaire fysica, Amsterdam, Netherlands
\\
$^{95}$Nuclear Physics Group, STFC Daresbury Laboratory, Daresbury, United Kingdom
\\
$^{96}$Nuclear Physics Institute, Academy of Sciences of the Czech Republic, \v{R}e\v{z} u Prahy, Czech Republic
\\
$^{97}$Oak Ridge National Laboratory, Oak Ridge, Tennessee, United States
\\
$^{98}$Petersburg Nuclear Physics Institute, Gatchina, Russia
\\
$^{99}$Physics Department, Creighton University, Omaha, Nebraska, United States
\\
$^{100}$Physics department, Faculty of science, University of Zagreb, Zagreb, Croatia
\\
$^{101}$Physics Department, Panjab University, Chandigarh, India
\\
$^{102}$Physics Department, University of Cape Town, Cape Town, South Africa
\\
$^{103}$Physics Department, University of Jammu, Jammu, India
\\
$^{104}$Physics Department, University of Rajasthan, Jaipur, India
\\
$^{105}$Physikalisches Institut, Eberhard Karls Universit\"{a}t T\"{u}bingen, T\"{u}bingen, Germany
\\
$^{106}$Physikalisches Institut, Ruprecht-Karls-Universit\"{a}t Heidelberg, Heidelberg, Germany
\\
$^{107}$Physik Department, Technische Universit\"{a}t M\"{u}nchen, Munich, Germany
\\
$^{108}$Purdue University, West Lafayette, Indiana, United States
\\
$^{109}$Research Division and ExtreMe Matter Institute EMMI, GSI Helmholtzzentrum f\"ur Schwerionenforschung GmbH, Darmstadt, Germany
\\
$^{110}$Rudjer Bo\v{s}kovi\'{c} Institute, Zagreb, Croatia
\\
$^{111}$Russian Federal Nuclear Center (VNIIEF), Sarov, Russia
\\
$^{112}$Saha Institute of Nuclear Physics, Kolkata, India
\\
$^{113}$School of Physics and Astronomy, University of Birmingham, Birmingham, United Kingdom
\\
$^{114}$Secci\'{o}n F\'{\i}sica, Departamento de Ciencias, Pontificia Universidad Cat\'{o}lica del Per\'{u}, Lima, Peru
\\
$^{115}$SSC IHEP of NRC Kurchatov institute, Protvino, Russia
\\
$^{116}$Stefan Meyer Institut f\"{u}r Subatomare Physik (SMI), Vienna, Austria
\\
$^{117}$SUBATECH, IMT Atlantique, Universit\'{e} de Nantes, CNRS-IN2P3, Nantes, France
\\
$^{118}$Suranaree University of Technology, Nakhon Ratchasima, Thailand
\\
$^{119}$Technical University of Ko\v{s}ice, Ko\v{s}ice, Slovakia
\\
$^{120}$Technical University of Split FESB, Split, Croatia
\\
$^{121}$The Henryk Niewodniczanski Institute of Nuclear Physics, Polish Academy of Sciences, Cracow, Poland
\\
$^{122}$The University of Texas at Austin, Physics Department, Austin, Texas, United States
\\
$^{123}$Universidad Aut\'{o}noma de Sinaloa, Culiac\'{a}n, Mexico
\\
$^{124}$Universidade de S\~{a}o Paulo (USP), S\~{a}o Paulo, Brazil
\\
$^{125}$Universidade Estadual de Campinas (UNICAMP), Campinas, Brazil
\\
$^{126}$Universidade Federal do ABC, Santo Andre, Brazil
\\
$^{127}$University of Houston, Houston, Texas, United States
\\
$^{128}$University of Jyv\"{a}skyl\"{a}, Jyv\"{a}skyl\"{a}, Finland
\\
$^{129}$University of Liverpool, Liverpool, United Kingdom
\\
$^{130}$University of Tennessee, Knoxville, Tennessee, United States
\\
$^{131}$University of the Witwatersrand, Johannesburg, South Africa
\\
$^{132}$University of Tokyo, Tokyo, Japan
\\
$^{133}$University of Tsukuba, Tsukuba, Japan
\\
$^{134}$Universit\'{e} de Lyon, Universit\'{e} Lyon 1, CNRS/IN2P3, IPN-Lyon, Villeurbanne, Lyon, France
\\
$^{135}$Universit\'{e} de Strasbourg, CNRS, IPHC UMR 7178, F-67000 Strasbourg, France, Strasbourg, France
\\
$^{136}$Universit\`{a} degli Studi di Pavia, Pavia, Italy
\\
$^{137}$Universit\`{a} di Brescia, Brescia, Italy
\\
$^{138}$V.~Fock Institute for Physics, St. Petersburg State University, St. Petersburg, Russia
\\
$^{139}$Variable Energy Cyclotron Centre, Kolkata, India
\\
$^{140}$Warsaw University of Technology, Warsaw, Poland
\\
$^{141}$Wayne State University, Detroit, Michigan, United States
\\
$^{142}$Wigner Research Centre for Physics, Hungarian Academy of Sciences, Budapest, Hungary
\\
$^{143}$Yale University, New Haven, Connecticut, United States
\\
$^{144}$Yonsei University, Seoul, Republic of Korea
\\
$^{145}$Zentrum f\"{u}r Technologietransfer und Telekommunikation (ZTT), Fachhochschule Worms, Worms, Germany
\endgroup